\documentclass[useAMS,usegraphicx,usenatbib,onecolumn]{mn2e}
\usepackage{amssymb}
\usepackage{times}
\newcounter{subequation}
\newcommand{\D}[2]{\frac{\partial #2}{\partial #1}}
\newcommand\bb[1] {\mbox{\boldmath{$#1$}}}
\newcommand\del{\bb{\nabla}}
\newcommand\bcdot{\,\bb{\cdot}\,}
\newcommand\btimes{\,\bb{\times}\,}
\newcommand{\mc}[1]{\mathcal{#1}}
\newcommand{\msb}[1]{\bb{\mathsf{#1}}}
\newcommand{\valf}{\bb{v_{\rm A}}}
\newcommand{\kdotva}{(\bb{k}\bcdot\valf)^2}
\newcommand{\vasq}{v^2_{\rm A}}
\newcommand{\kdotb}{\bb{k}\bcdot\bb{B}}
\newcommand{\kxva}{\bb{k}\btimes\valf}
\newcommand{\rhoi}{\rho_{\rm i}}
\newcommand{\nem}{n_{\rm e}}
\newcommand{\ex}{\hat{\bb{e}}_x}
\newcommand{\ey}{\hat{\bb{e}}_y}
\newcommand{\ez}{\hat{\bb{e}}_z}
\newcommand{\hall}{\frac{ck_z(\kdotb)}{4\pi en_{\rm e}}}

\title[Shear Instabilities in Weakly-Ionized Flows]{On the Linear Stability of Weakly-Ionized, Magnetized Planar Shear Flows}
\author[M. W. Kunz]{Matthew W. Kunz\thanks{Email: mkunz@uiuc.edu}\\
Department of Physics, University of Illinois at Urbana-Champaign,
1002 W. Green Street, Urbana, IL 61801}
\date{Released 2007}
\pagerange{\pageref{firstpage}--\pageref{lastpage}} \pubyear{2007}
\def\LaTeX{L\kern-.36em\raise.3ex\hbox{a}\kern-.15em
    T\kern-.1667em\lower.7ex\hbox{E}\kern-.125emX}
\begin{document}
\label{firstpage} \maketitle
\begin{abstract}
We investigate the effects of ambipolar diffusion and the Hall
effect on the stability of weakly-ionized, magnetized planar shear
flows. Employing a local approach similar to the shearing-sheet
approximation, we solve for the evolution of linear perturbations
in both streamwise-symmetric and non-streamwise-symmetric
geometries using WKB techniques and/or numerical methods. We find
that instability arises from the combination of shear and
non-ideal magnetohydrodynamic processes, and is a result of the
ability of these processes to influence the free energy path
between the perturbations and the shear. They turn what would be
simple linear-in-time growth due to current and vortex stretching
from shear into exponentially-growing instabilities. Our results
aid in understanding previous work on the behaviour of
weakly-ionized accretion discs. In particular, the recent finding
that the Hall effect and ambipolar diffusion destabilize both
positive and negative angular velocity gradients acquires a
natural explanation in the more general context of this paper. We
construct a simple toy model for these instabilities based upon
transformation operators (shears, rotations, and projections) that
captures both their qualitative and, in certain cases, exact
quantitative behaviour.
\end{abstract}


\section{Introduction}

The importance of understanding the physics of shear flows has
been appreciated for well over a century, starting with the
pioneering work of \citet{helmholtz1868} and \citet{kelvin1871}.
The stability of these flows in a wide variety of situations has
been thoroughly studied and reviewed, most notably in
Chandrasekhar's (1961) classic text. Since that time, observations
have revealed that shear flows are commonplace in astrophysical
systems, playing likely roles in the stability and collimation of
jets \citep{ftz81,fj84,bbr84}, the solar corona \citep{kads93},
bipolar outflows from young stellar objects
\citep{pringle89,bachiller96}, rotating stars
\citep{cowling51,gs67}, accretion discs \citep{pl95,bh98}, and
even the Earth's magnetopause \citep{mckenzie70}. In nearly all of
these systems, magnetic fields play a vital role in determining
the structure and evolution. There are, however, systems where the
importance of magnetic fields remain unclear, due to poor
ionization and therefore weak coupling between the dominant ion
and neutral species and the magnetic field. These include
molecular clouds and their cores \citep{mc99}, galactic molecular
discs \citep{bb94}, protostellar accretion discs
\citep{gammie96,hs98,ftb02}, protostellar outflows and disc winds
\citep{wk93}, dwarf nova disks \citep{gm98,ss03}, shock waves in
dense molecular clouds \citep{wardle91,dmk93,rc07}, and
protoplanetary discs \citep{sm99,smun00,sw05,cmc07}. In these
systems, various non-ideal magnetohydrodynamic (MHD) effects may
come into play. For example, the neutral particles may drift
relative to the ions in a process referred to as ambipolar
diffusion. If, on the other hand, the ions drift relative to the
electrons (and thereby the magnetic field), then the Hall effect
occurs.

Given the high occurrence of shear flows and low levels of
ionization in various astrophysical environments, it is not
surprising that examples can be found when the two coincide.
Perhaps the most obvious examples are outflows from star-forming
regions, where shear instabilities can occur at the interface
between the jet and the ambient material \citep{wzhc04}. The
resulting turbulent boundary layer can transfer linear momentum
from the jet to the ambient medium, suggesting a possible
mechanism for entrainment. Even at the launching point of these
outflows, there may be differing velocity profiles in the ion and
neutral fluids (see fig. 2 of Wardle \& K\"{o}nigl 1993),
conditions ripe for both ambipolar diffusion and shear. Another
notable example concerns the interface between the
magnetically-active and -inactive (`dead') regions in
protoplanetary discs. It has been suggested that the well-coupled
outer layers of protoplanetary discs will be subject to the
magnetorotational instability (MRI; Balbus \& Hawley 1991), while
the shielded midplane layers will remain dormant \citep{gammie96}.
In this situation, angular momentum can be effectively transported
in the outer layers but not in the midplane, setting up a velocity
profile in the vertical (away from the midplane) direction
\citep{fs03,fn06}.

Here we undertake a study of a magnetized, planar shear layer in
the presence of ambipolar diffusion and the Hall effect, and find
that instability arises when shear and non-ideal MHD effects act
in concert. The instabilities are similar to those found by
\citet{wardle99}, \citet{bt01}, \citet{kb04}, and \citet{desch04}
in the context of the MRI. Here we show that they are actually
much more general. Our approach not only provides us with a
clearer physical picture of the instabilities, free from the
complications of rotation, but also brings qualitatively new
results. In the case of ambipolar diffusion, instability arises
from the combination of shear and the anisotropic nature of its
wave damping. When the Hall effect is present, epicyclic-like
motions set up by electromagnetic (whistler) waves couple to the
background shear. When the handedness of these waves is opposite
to that of the shear, instability occurs. It is notable that
neither of these processes relies on rotational kinematics. In
fact, these instabilities are neither versions of the
Kelvin-Helmholtz instability \citep{wzhc04}, nor versions of the
MRI \citep{bh91,bh92b}, despite their reliance on the presence of
both shear and magnetic fields. The unstable modes are a result of
the ability of non-ideal MHD processes to open new pathways for
the fluid to tap into the free energy of shear.

An outline of the paper is as follows. In \S \ref{S_equations}, we
discuss the formulation of the problem and give the basic
equations to be solved. We then consider the evolution of Eulerian
perturbations to these equations in comoving, local Lagrangian
coordinates \citep{glb65}. After deriving two coupled equations
for the evolution of the relevant magnetic field eigenvectors in
\S \ref{S_shearingsheet}, we solve them both analytically and
numerically in several different situations. In \S \ref{S_AD}, we
first restrict our attention to the effects of ambipolar diffusion
on the stability of the system. After a detailed analysis and
discussion, we then examine the effects of Hall electromotive
forces (HEMFs) in Section \ref{S_hall}. Contact with prior work is
emphasized. In \S \ref{S_discussion}, we construct and analyse a
toy model that captures all the salient features of these
instabilities. Section \ref{S_summary} summarizes our findings and
conclusions.

\section{Formulation of the Problem}

\subsection{Basic Equations}\label{S_equations}

The equations describing a non-ideal MHD system, in the limit of
negligible ion and electron inertia, are the continuity equation,
\renewcommand{\theequation}{\arabic{equation}\alph{subequation}}
\setcounter{subequation}{1}
\begin{equation}\label{E_continuity}
\D{t}{\rho} + \del\bcdot(\rho\bb{v}) = 0\,,
\end{equation}
the force equation,
\addtocounter{equation}{-1}\addtocounter{subequation}{1}
\begin{equation}\label{E_force}
\D{t}{\bb{v}} + \bb{v}\bcdot\del\bb{v} = -\frac{1}{\rho}\del P +
\frac{\bb{j}\btimes\bb{B}}{c\rho}\,,
\end{equation}
the magnetic induction equation,
\addtocounter{equation}{-1}\addtocounter{subequation}{1}
\begin{equation}\label{E_induction}
\D{t}{\bb{B}} = \del\btimes\left[\bb{v}\btimes\bb{B} -
\frac{\bb{j}\btimes\bb{B}}{e\nem} +
\frac{(\bb{j}\btimes\bb{B})\btimes\bb{B}}{c\gamma\rhoi\rho}\right]\,,
\end{equation}
and Amper\'{e}'s law,
\addtocounter{equation}{-1}\addtocounter{subequation}{1}
\begin{equation}\label{E_Ampere}
\bb{j} = \frac{c}{4\pi}\del\btimes\bb{B}\,.
\end{equation}
\renewcommand{\theequation}{\arabic{equation}}
Our notation is standard: $\rho$ is the mass density, $\bb{v}$ is
the velocity, $P$ is the gas pressure, $\bb{B}$ is the magnetic
field, and $\bb{j}$ is the current density. The density, velocity,
and pressure all refer to the dominant neutral species. The
combination $\gamma\rhoi$ is the neutral-ion collision frequency,
with
\[
\gamma = \frac{\langle\sigma w\rangle_{\rm in}}{m_{\rm i} + m_{\rm
n}}
\]
being the drag coefficient, and $\nem$ the electron number
density. The quantity $\langle \sigma w\rangle_{\rm in}$ is the
average collisional rate between ions of mass $m_{\rm i}$ and
neutrals of mass $m_{\rm n}$; it is equal to $1.69 \times
10^{-9}\,(T/10\,{\rm K})^{1/2}$ cm$^{3}$ s$^{-1}$ for
HCO$^+$-H$_2$ collisions, and is almost identical to this value
for Na$^+$-H$_2$ and Mg$^+$-H$_2$ collisions (see McDaniel \&
Mason 1973).

The three terms on the right-side of Equation (\ref{E_induction})
represent induction, the Hall effect, and ambipolar diffusion,
respectively. The difference between ion and electron velocities
gives rise to the Hall effect, whereas the difference between ion
and neutral velocities gives rise to ambipolar diffusion.
Discussions of their relative magnitudes can be found in
\citet{bt01} and \citet{ss02}. For the sake of completeness,
however, we repeat here the relative ratio of the ambipolar to
Hall terms:
\begin{equation}
\frac{{\rm Am}}{{\rm Ha}} \sim \left(\frac{10^{13}\,{\rm
cm}^{-3}}{n}\right)^{1/2}\,\left(\frac{T}{10^3\,{\rm
K}}\right)^{1/2}\,\left(\frac{v_{\rm A}}{c_{\rm s}}\right)\,,
\end{equation}
where $n$ is the number density of the neutrals and $c_{\rm s}$ is
the isothermal sound speed. Assuming that the final two factors
are each about 0.1, we see that a neutral density below about
$10^9$ cm$^{-3}$ brings us safely into the ambipolar diffusion
regime (see fig. 1 of Kunz \& Balbus 2004). Since we are concerned
here with the interplay between non-ideal MHD effects and shear,
we ignore Ohmic dissipation, for which no shear instabilities are
present. In writing Equation (\ref{E_induction}), we have
implicitly assumed that the thermal-pressure force on the ions and
electrons is negligible compared to the electromagnetic and
collisional forces. This is an excellent approximation for the
systems of interest. In addition, the inelastic momentum transfer
by the ion and electron fluids due to attachment onto grains and
neutralization is negligible compared to the momentum transfer due
to elastic collisions, and it has been implicitly omitted from the
induction equation. We have also ignored coupling via ionization
and recombination, since these processes are slow compared to
elastic processes.

Consider a flow in a stationary (`lab') Cartesian coordinate
system along the $y$-axis, $\bb{v} = v(x)\ey$. Although we take
the density and the magnetic field to be everywhere uniform, we
put no restrictions on the $x$-dependance of the velocity field
and the orientation of the magnetic field. Eulerian perturbations
to this flow are allowed, denoted by a $\delta$. Keeping only
terms linear in $\delta$ and working in the Boussinesq
approximation, we find
\renewcommand{\theequation}{\arabic{equation}\alph{subequation}}
\setcounter{subequation}{1}
\begin{equation}
\del\bcdot\delta\bb{v} = 0\,,
\end{equation}
\addtocounter{equation}{-1}\addtocounter{subequation}{1}
\begin{equation}
\D{t}{\delta\bb{v}} + v(x)\D{y}{\delta\bb{v}} + \ey\delta
v_x\frac{dv(x)}{dx} = -\del\left(\frac{\delta P}{\rho} +
\frac{\bb{B}\bcdot\delta\bb{B}}{4\pi\rho}\right) +
\frac{1}{4\pi\rho}(\bb{B}\bcdot\del) \delta\bb{B}\,,
\end{equation}
\addtocounter{equation}{-1}\addtocounter{subequation}{1}
\begin{equation}
\D{t}{\delta\bb{B}} + v(x)\D{y}{\delta\bb{B}} - \ey \delta B_x
\frac{dv(x)}{dx} = (\bb{B}\bcdot\del)\delta\bb{v} -
\left(\frac{c\bb{B}\bcdot\del}{4\pi
e\nem}\right)\,(\del\btimes\delta\bb{B}) +
\left(\frac{\valf\btimes\del}{\gamma\rhoi}\right)\del\bcdot(\valf\btimes\delta\bb{B})
+ \frac{v^2_{\rm A}}{\gamma\rhoi}\del^2\delta\bb{B}\,,
\end{equation}
\renewcommand{\theequation}{\arabic{equation}}
where
\[
\valf = \frac{\bb{B}}{(4\pi\rho)^{1/2}}
\]
is the Alfv\'{e}n velocity.

\subsection{Shearing Sheet Formalism}\label{S_shearingsheet}

The equations are first transformed from our lab-frame coordinate
system to one comoving with the flow, centered at a fiducial
location $(x_0,y_0,z_0)$ moving at velocity $v(x_0)=v_0$. We then
consider a local neighborhood surrounding this point and Taylor
expand the velocity field about $x_0$ to find
\[
v(x) = v_0 + (x-x_0) \left.\frac{dv}{dx}\right|_{x_0}\,.
\]
As is well known, a shearing background precludes simple plane
wave solutions to the perturbation equations given in the previous
section \citep{glb65}. This difficulty is circumvented by adopting
shearing coordinates, given by
\renewcommand{\theequation}{\arabic{equation}\alph{subequation}}
\setcounter{subequation}{1}
\begin{equation}
x' = x\,,
\end{equation}
\addtocounter{equation}{-1}\addtocounter{subequation}{1}
\begin{equation}
y' = y - 2Axt\,,
\end{equation}
\addtocounter{equation}{-1}\addtocounter{subequation}{1}
\begin{equation}
z' = z\,,
\end{equation}
\addtocounter{equation}{-1}\addtocounter{subequation}{1}
\begin{equation}
t' = t\,,
\end{equation}
\renewcommand{\theequation}{\arabic{equation}}
so that
\renewcommand{\theequation}{\arabic{equation}\alph{subequation}}
\setcounter{subequation}{1}
\begin{equation}
\D{x}{} = \D{x'}{} - 2At'\D{y'}{}\,,
\end{equation}
\addtocounter{equation}{-1}\addtocounter{subequation}{1}
\begin{equation}
\D{y}{} = \D{y'}{}\,,
\end{equation}
\addtocounter{equation}{-1}\addtocounter{subequation}{1}
\begin{equation}
\D{z}{} = \D{z'}{}\,,
\end{equation}
\addtocounter{equation}{-1}\addtocounter{subequation}{1}
\begin{equation}
\D{t}{} = \D{t'}{} - 2Ax'\D{y'}{}\,,
\end{equation}
\renewcommand{\theequation}{\arabic{equation}}
where we have defined $2A \equiv (dv/dx)_{x_0}$.\footnote{The
parameter $2A$ used in this paper is not to be confused with the
Oort $A$ constant, which concerns rotating systems. Here, we are
primarily interested in planar shear flow, and $2A$ is to be
identified as the characteristic frequency associated with the
velocity profile.} In this frame, the velocity field is $v(x') =
2Ax'$. The benefit of this coordinate transformation is that a
spatial dependence $\exp[i(k'_x x' + k'_y y' + k'_z z')]$ may be
assumed for the perturbations, so long as we replace a fixed $x$
wavenumber with a shearing one:
\begin{equation}\label{E_kx}
k_x \leftarrow k_x(t) = k'_x - 2Ak'_y t'\,.
\end{equation}
No modification of $y'$ and $z'$ variables are needed. Enacting
this transformation and dropping the primes for ease of notation,
our equations become
\renewcommand{\theequation}{\arabic{equation}\alph{subequation}}
\setcounter{subequation}{1}
\begin{equation}\label{E_cont}
k_x \delta v_x + k_y \delta v_y + k_z \delta v_z = 0\,,
\end{equation}
\addtocounter{equation}{-1}\addtocounter{subequation}{1}
\begin{equation}\label{E_vx}
\frac{d\delta v_x}{dt} + ik_x \left(\frac{\delta P}{\rho} +
\frac{\bb{B}\bcdot\delta\bb{B}}{4\pi\rho}\right) -
i\frac{\kdotb}{4\pi\rho}\,\delta B_x=0\,,
\end{equation}
\addtocounter{equation}{-1}\addtocounter{subequation}{1}
\begin{equation}\label{E_vy}
\frac{d\delta v_y}{dt} + 2A\delta v_x + ik_y \left(\frac{\delta
P}{\rho} + \frac{\bb{B}\bcdot\delta\bb{B}}{4\pi\rho}\right) -
i\frac{\kdotb}{4\pi\rho}\,\delta B_y=0\,,
\end{equation}
\addtocounter{equation}{-1}\addtocounter{subequation}{1}
\begin{equation}\label{E_vz}
\frac{d\delta v_z}{dt} + ik_z \left(\frac{\delta P}{\rho} +
\frac{\bb{B}\bcdot\delta\bb{B}}{4\pi\rho}\right)
-i\frac{\kdotb}{4\pi\rho}\,\delta B_z = 0\,,
\end{equation}
\addtocounter{equation}{-1}\addtocounter{subequation}{1}
\begin{equation}\label{E_bx}
\frac{d\delta B_x}{dt} - i(\kdotb) \delta v_x -
\frac{c(\kdotb)}{4\pi e\nem}\,(\bb{k}\btimes\delta\bb{B})_x +
\frac{k^2 v^2_A}{\gamma\rhoi}\,\delta B_x -
\frac{(\kxva)_x}{\gamma\rhoi}\,(\kxva)\bcdot\delta\bb{B} = 0\,,
\end{equation}
\addtocounter{equation}{-1}\addtocounter{subequation}{1}
\begin{equation}\label{E_by}
\frac{d\delta B_y}{dt} - 2A\delta B_x - i(\kdotb) \delta v_y -
\frac{c(\kdotb)}{4\pi e\nem}\,(\bb{k}\btimes\delta\bb{B})_y +
\frac{k^2v^2_A}{\gamma\rhoi}\,\delta B_y -
\frac{(\kxva)_y}{\gamma\rhoi}\,(\kxva)\bcdot\delta\bb{B} = 0\,,
\end{equation}
\addtocounter{equation}{-1}\addtocounter{subequation}{1}
\begin{equation}\label{E_bz}
\frac{d\delta B_z}{dt} - i(\kdotb) \delta v_z -
\frac{c(\kdotb)}{4\pi e\nem}\,(\bb{k}\btimes\delta\bb{B})_z
+\frac{k^2v^2_A}{\gamma\rhoi}\,\delta B_z -
\frac{(\kxva)_z}{\gamma\rhoi}\,(\kxva)\bcdot\delta\bb{B} = 0\,,
\end{equation}
\renewcommand{\theequation}{\arabic{equation}}
where the perturbations are now time-dependent Fourier amplitudes
and we have suppressed the explicit time-dependent notation in
$k_x$ and $k^2\equiv k^2_x + k^2_y + k^2_z$. Since $dk_x/dt =
-2Ak_y$, Equations (\ref{E_bx}) - (\ref{E_bz}) together with
Equation (\ref{E_cont}) guarantee the divergence free condition
$d(\bb{k}\bcdot\delta\bb{B})/dt=0$.

In an attempt to keep the presentation as simple as possible, we
first consider Equations (7) in the limit where ambipolar
diffusion is the dominant non-ideal MHD process (\S \ref{S_AD}),
ignoring the Hall effect for the time being. Not only does this
aid in our interpretation of the physics, but also the two
processes generally act in distinct regions of parameter space. We
then isolate the Hall effect in \S \ref{S_hall}, neglecting
ambipolar diffusion. The similarities and differences of the two
resulting instabilities are discussed in Sections
\ref{S_discussion} and \ref{S_summary}. Much of the formalism for
understanding the Hall--shear instability is developed in the
following section on ambipolar diffusion, and so it behooves us to
encourage any readers primarily interested in Hall physics not to
bypass the following section.

\section{Ambipolar-Diffusion--Shear Instability}\label{S_AD}

In this Section, we are concerned solely with the interplay of
ambipolar diffusion and shear, and we neglect the Hall terms in
Equations (\ref{E_bx})-(\ref{E_bz}). Before we begin reducing
these equations to a more manageable set, however, it is of
interest to note that, while the presence of shear causes the
$y$-component of the background magnetic field to grow linearly
with time:
\[
B_y(t) = B_y(0) + 2AB_xt\,,
\]
where $B_y(0)$ is the initial $y$ field, the combination
$\bb{k}\bcdot\bb{B}$ is constant with time, despite the fact that
neither the Eulerian wavenumber $\bb{k}=[k_x(t),k_y,k_z]$ nor the
magnetic field vector $\bb{B}$ is individually constant:
\[
\bb{k}\bcdot\bb{B} = k'_x B_x + k_y B_y(0) + k_z B_z\,.
\]
Unfortunately, the same does not hold for the combination
$\bb{k}\btimes\bb{B}$, and so the ambipolar diffusion terms in
Equations (\ref{E_bx})-(\ref{E_bz}) are intrinsically
time-dependent. This complicates matters.

We simplify the set of Equations (7) as follows. Using
$\bb{k}\bcdot\delta\bb{B}=0$, we first obtain
\begin{equation}\label{E_bz2}
\delta B_z = -\frac{1}{k_z}\,\bigl(k_x\delta B_x + k_y\delta
B_y\bigr)\,.
\end{equation}
Then we may eliminate $\delta B_z$ from Equations (\ref{E_bx}) and
(\ref{E_by}) to find
\renewcommand{\theequation}{\arabic{equation}\alph{subequation}}
\setcounter{subequation}{1}
\begin{equation}\label{E_vx2}
\delta v_x = \frac{1}{i(\kdotb)}\left[\left(\frac{d}{dt} +
k^2\eta_{xx}\right)\delta B_x + k^2\eta_{xy}\delta B_y\right]\,,
\end{equation}
\addtocounter{equation}{-1}\addtocounter{subequation}{1}
\begin{equation}\label{E_vy2}
\delta v_y = \frac{1}{i(\kdotb)}\left[\left(\frac{d}{dt} +
k^2\eta_{yy}\right)\delta B_y + \bigl(k^2\eta_{yx} -
2A\bigr)\delta B_x\right]\,.
\end{equation}
\renewcommand{\theequation}{\arabic{equation}}
Here we have introduced the resistivity tensor $\bb{\eta}$, whose
elements are given by
\begin{equation}\label{E_eta}
\eta_{ij} = \frac{\vasq}{\gamma\rhoi}\,\delta_{ij} -
\frac{(\hat{\bb{k}}\btimes\valf)_i\,(\hat{\bb{k}}\btimes\valf)_j}{\gamma\rhoi}
+\frac{k_j}{k_z}\frac{(\hat{\bb{k}}\btimes\valf)_i\,(\hat{\bb{k}}\btimes\valf)_z}{\gamma\rhoi}\,,
\end{equation}
where $\delta_{ij}$ is the usual Kronecker delta function and
$\hat{\bb{k}}$ is the unit wavevector. Next, we rearrange Equation
(\ref{E_vy}):
\begin{equation}\label{E_dp}
\frac{\delta P}{\rho} + \frac{\bb{B}\bcdot\delta\bb{B}}{4\pi\rho}
= \frac{1}{k_y(\kdotb)}\left[\frac{d^2\delta B_y}{dt^2} +
\frac{d}{dt}\bigl(k^2\eta_{yy}\delta B_y + k^2\eta_{yx}\delta
B_x\bigr) + \kdotva\delta B_y + 2A\bigl(k^2\eta_{xy}\delta B_y +
k^2\eta_{xx}\delta B_x\bigr)\right]\,.
\end{equation}
Inserting Equations (\ref{E_vx2}) and (\ref{E_dp}) into Equation
(\ref{E_vx}), we obtain
\begin{eqnarray}\label{E_main1}
\lefteqn{\frac{d^2\delta B_x}{dt^2} +
\frac{d}{dt}\left(k^2\eta_{xx}\delta
B_x-\frac{k_x}{k_y}k^2\eta_{yx}\delta B_x\right) + \kdotva\delta
B_x -
2A\frac{k_x}{k_y}\left(k^2\eta_{xx}+\frac{k_y}{k_x}k^2\eta_{yx}\right)\delta
B_x}\nonumber\\*&&\mbox{}=\frac{k_x}{k_y}\left[\frac{d^2\delta
B_y}{dt^2} + \frac{d}{dt}\left(k^2\eta_{yy}\delta
B_y-\frac{k_y}{k_x}k^2\eta_{xy}\delta B_y\right) + \kdotva\delta
B_y + 2A\frac{k^2_{xy}}{k^2_x}k^2\eta_{xy}\delta B_y\right]\,.
\end{eqnarray}
For economy of notation, we have defined $k^2_{xy} \equiv k^2_x +
k^2_y$. We need another independent differential equation coupling
$\delta B_x$ and $\delta B_y$. Multiplying Equation (\ref{E_vy})
by $k_z$ and Equation (\ref{E_vz}) by $k_y$, then subtracting one
from the other, we find
\begin{equation}\label{E_vort}
k_z\frac{d\delta v_y}{dt} + 2Ak_z\delta v_x - k_y\frac{d\delta
v_z}{dt} - ik_z\frac{\kdotb}{4\pi\rho}\,\delta B_y +
ik_y\frac{\kdotb}{4\pi\rho}\,\delta B_z = 0\,.
\end{equation}
Using Equations (\ref{E_cont}), (\ref{E_bz2}), and (\ref{E_vy2})
in Equation (\ref{E_vort}) leads after some simplification to
\begin{eqnarray}\label{E_main2}
\frac{d^2\delta B_x}{dt^2} + \frac{d}{dt}\left(k^2\eta_{xx}\delta
B_x + \frac{k^2_{yz}}{k_xk_y}k^2\eta_{yx}\delta B_x\right)
-4A\frac{k_y}{k_x}\frac{d\delta B_x}{dt} + \kdotva\delta B_x +
2A\left(\frac{k^2_z-k^2_y}{k_xk_y}k^2\eta_{xx} -
\frac{k^2_{yz}}{k^2_x}k^2\eta_{yx}\right)\delta B_x
\nonumber\\*\mbox{} =
-\frac{k^2_{yz}}{k_xk_y}\left[\frac{d^2\delta B_y}{dt^2} +
\frac{d}{dt}\left(k^2\eta_{yy}\delta B_y +
\frac{k_xk_y}{k^2_{yz}}k^2\eta_{xy}\delta B_y\right) +
\kdotva\delta B_y + 2A\frac{k^2_z}{k^2_{yz}}k^2\eta_{xy}\delta
B_y\right]\,.
\end{eqnarray}
Again, for economy of notation, we have defined $k^2_{yz}\equiv
k^2_y + k^2_z$.

For numerical work, it is convenient to isolate the second-order
time derivatives. Equations (\ref{E_main1}) and (\ref{E_main2})
may be recombined to yield
\begin{equation}\label{E_1}
\frac{d^2\delta B_x}{dt^2} =-\frac{d}{dt}\bigl(k^2\eta_{xx}\delta
B_x + k^2\eta_{xy}\delta B_y\bigr) +
4A\frac{k_xk_y}{k^2}\frac{d\delta B_x}{dt} - \kdotva\delta B_x +
4A\frac{k_xk_y}{k^2}\bigl(k^2\eta_{xx}\delta B_x +
k^2\eta_{xy}\delta B_y\bigr)\,,
\end{equation}
\begin{equation}\label{E_2}
\frac{d^2\delta B_y}{dt^2} = -\frac{d}{dt}\bigl(k^2\eta_{yy}\delta
B_y + k^2\eta_{yx}\delta B_x\bigr) +
4A\frac{k^2_y}{k^2}\frac{d\delta B_x}{dt} - \kdotva\delta B_y -
2A\frac{k^2_{xz}-k^2_y}{k^2}\bigl(k^2\eta_{xx}\delta B_x +
k^2\eta_{xy}\delta B_y\bigr)\,.
\end{equation}
Equations (\ref{E_1}) and (\ref{E_2}) are the two coupled
differential equations in $\delta B_x$ and $\delta B_y$ that form
the cornerstone of the analysis.

\subsection{Qualitative Behavior}\label{S_behaviour}

In a system where ambipolar diffusion and shear are absent, it is
straightforward to show from Equations (\ref{E_1}) and (\ref{E_2})
that the perturbed magnetic field lines simply follow fluid
elements, resulting in Doppler-shifted Alfv\'{e}n waves
propagating along the background magnetic field. The introduction
of shear into the picture has two effects. The first effect is
that vorticity is generated in the flow. The accompanying
centrifugal force (associated with the resulting eddy) pushes on
the shear interface, resulting in the growth of any deformation in
the interface with time, provided that the vorticity is out of
phase with the surface deformation. This is the essence of the
Kelvin-Helmholtz instability. The second effect is that any
$x$-displacement in the magnetic field becomes sheared out into an
$y$-displacement. Thus, the perturbed magnetic field is
effectively rotated until it becomes aligned with the shear
interface.

In the presence of shear, any physical mechanism that conspires to
rotate $\delta B_y$ {\em back} into $\delta B_x$ completes a
feedback loop and results in growth. Ambipolar diffusion does just
that. Since ambipolar diffusion only affects those currents
flowing perpendicular to the background magnetic field, it tends
to align magnetic field perturbations perpendicular to the
background magnetic field. This manifests itself as an effective
rotation of $\delta B_y$ into $\delta B_x$ (albeit with a decrease
in $|\delta\bb{B}|$). In this case, ambipolar diffusion and shear
conspire to stretch any perturbation in the magnetic field,
resulting in an exponentially-growing instability. In a gas where
either (1) the magnetic field is so strong that its tension
effectively resists being stretched by the shear [i.e., $\kdotva
\gg 4A^2$], or (2) the bulk neutral fluid is so poorly coupled to
the magnetic field so that its velocity does not grow with it
[i.e., $(\gamma\rhoi) \ll 2A$], this instability does not operate
efficiently. We will show, however, that there still remains a
great deal of unstable parameter space with which to work.

This route to instability has been seen before in the
ambipolar-diffusion--modified MRI \citep{kb04,desch04}, where the
role of shear is played by the differential rotation of an
accretion disc [$v(x) = x\Omega(x)$, where $\Omega$ is the orbital
frequency]. Here, however, we see that this destabilizing
behaviour is part of a more general process, and in no way depends
on rotational kinematics. The finding that ambipolar diffusion
renders an accretion disc unstable (albeit weakly) for both
inwardly- {\em and} outwardly-decreasing angular velocity profiles
acquires a natural explanation in the more general context of this
paper. Ambipolar diffusion can destabilize {\em any} shear flow
profile, in very much the same way that the sign of $dv(x)/dx$
does not determine the outcome of the Kelvin-Helmholtz
instability. In fact, we will show that the criterion for this
ambipolar-diffusion shear instability is independent of the
magnetic field strength, and is reliant only upon the ratio of the
ion-neutral collision frequency $\gamma\rhoi$ to the frequency
implied by the shear of the flow, $2A$, and the geometry of the
magnetic field.

\subsection{The Case $k_y=B_x=0$: a Time-Independent Zero-Order
State}\label{S_case1}

The qualitative behaviour discussed above is most easily seen in
the simple case of $k_y=B_x=0$. In this situation, the zero-order
state is time-independent, and we may seek solutions to Equations
(\ref{E_1}) and (\ref{E_2}) with time dependence $\exp(\sigma t)$.
The resulting dispersion relation is
\begin{equation}\label{E_disprel}
\sigma^4 + \left[\frac{k^2\vasq +
\kdotva}{\gamma\rhoi}\right]\sigma^3 + \mc{C}_2\sigma^2 +
\kdotva\left[\frac{k^2\vasq + \kdotva}{\gamma\rhoi}\right]\sigma +
\mc{C}_0 = 0\,,
\end{equation}
where
\renewcommand{\theequation}{\arabic{equation}\alph{subequation}}
\setcounter{subequation}{1}
\begin{equation}
\mc{C}_2 = 2\kdotva + \frac{k^2\vasq\kdotva}{(\gamma\rhoi)^2} -
2A\,\frac{k_x B_y (\kdotb)}{4\pi\gamma\rhoi\rho}\,,
\end{equation}
\addtocounter{equation}{-1}\addtocounter{subequation}{1}
\begin{equation}
\mc{C}_0 = \kdotva\left[\kdotva -
2A\,\frac{k_xB_y(\kdotb)}{4\pi\gamma\rhoi\rho}\right]\,.
\end{equation}
\renewcommand{\theequation}{\arabic{equation}}
Before proceeding to obtain an instability criterion, let us note
that Equation (\ref{E_disprel}) may be written in the more compact
form
\begin{equation}\label{E_wavedamping}
\left[\sigma^2 + \frac{k^2\vasq}{\gamma\rhoi}\,\sigma +
\kdotva\right]\,\left[\sigma^2 +
\frac{\kdotva}{\gamma\rhoi}\,\sigma + \kdotva\right] =
2A\,\frac{k_xB_y(\kdotb)}{4\pi\gamma\rhoi\rho}\bigl[\sigma^2 +
\kdotva\bigr]\,.
\end{equation}
Consider the limit of vanishing shear. In this case, the
right-hand side goes to zero and the two brackets on the left-hand
side become decoupled from one another. The two solutions obtained
by setting the left bracket to zero correspond to forward- and
backward-propagating Alfv\'{e}n waves with
$\delta\bb{B}\,||\,\bb{B}$, which are damped at a rate
$k^2\vasq/2\gamma\rhoi$. The other two solutions obtained by
setting the right bracket to zero correspond to forward- and
backward-propagating Alfv\'{e}n waves with $\delta\bb{B}\,||\,
(\hat{\bb{k}}\btimes\bb{B})$, which are damped at a rate
$\kdotva/2\gamma\rhoi$. One consequence of the difference in these
damping rates is an effective rotation of $\delta
\bb{B}\bcdot\bb{B}$ into
$\delta\bb{B}\bcdot(\hat{\bb{k}}\btimes\bb{B})$. As discussed in
\S \ref{S_behaviour}, this difference is at the heart of the
ambipolar-diffusion--shear instability.

A sufficient condition for unstable solutions to exist in Equation
(\ref{E_disprel}) is $\mc{C}_0 < 0$, or
\begin{equation}\label{E_crit1}
\kdotva - 2A\,\frac{k_xB_y(\kdotb)}{4\pi\gamma\rhoi\rho} < 0\,.
\end{equation}
This is identical to the result in \S 4.2 of \citet{kb04} in the
limit of vanishing rotation frequency $\Omega$ (so that the
epicyclic frequency $\kappa\rightarrow 0$), and is similar to the
instability criterion for the ideal MRI \citep{bh91}:
\[
\kdotva + \frac{k^2_z}{k^2}\frac{d\Omega^2}{d\ln R} < 0\,.
\]
Notice, however, that the stabilizing effects of ambipolar
diffusion seen in the second term of equation (35) of \citet{kb04}
[which is $\propto \kappa^2\kdotva/(\gamma\rhoi)^2$] are absent
here. This term represents epicyclic oscillations in the neutral
fluid, an inherently stabilizing motion in Keplerian discs, being
communicated to the (potentially) unstable ions through
collisional coupling. In the limit $\gamma\rhoi\rightarrow 0$
(i.e., infinite neutral-ion collision time-scale), it is clear
that this term dominates and the bulk neutral fluid oscillates at
the epicyclic frequency, unaffected by the presence of the
magnetic field and the ions that are tied to it. In other words,
if the neutrals can respond to magnetic forces on an epicyclic
time-scale, the MRI will act on both fluids and will be effective
in transporting angular momentum. Since it is rotation that gives
birth to these stabilizing epicyclic oscillations, this term is
absent in the dispersion relation given here. Evidentally, the
situation investigated in both \citet{kb04} and \citet{desch04}
was the superposition of two different, but related, instabilities
acting in tandem: the MRI, whose magnetic `tether' between fluid
elements, essential to the transport of angular momentum, is
undermined by the imperfect coupling between the ions and neutrals
(leading to decreased growth rates); and an
ambipolar-diffusion--shear instability, which results from the
combination of shear and the anisotropic damping of ambipolar
diffusion.

Further insight into the interpretation of Equation
(\ref{E_crit1}) is afforded by rewriting it in the form
\begin{equation}\label{E_crit2}
\frac{2A}{\gamma\rhoi} > \frac{k_zB_z}{k_xB_y}\,.
\end{equation}
Note that the strength of the magnetic field is not at all
relevant here; only the ratio of the shearing time-scale to the
neutral-ion collision time-scale and the geometry of the magnetic
field come into play. Furthermore, the freedom in choosing the
sign of $k_z/k_x$ guarantees that any non-constant velocity
profile can be destabilized, regardless of the sign of its
derivative. Physically, this equation states that the time for a
neutral particle to collide with an ion must be longer (by at
least the factor given on the right-hand side) than the time it
takes for a magnetic perturbation to grow by shear. If this
condition is not met, the neutral fluid is well-coupled to the
magnetic field, and we are left with simple linear-in-time growth
due to shearing of the magnetic field perturbation.

\begin{figure}
\begin{center}
\leavevmode \qquad\qquad\qquad\qquad\quad
\includegraphics[width=3in]{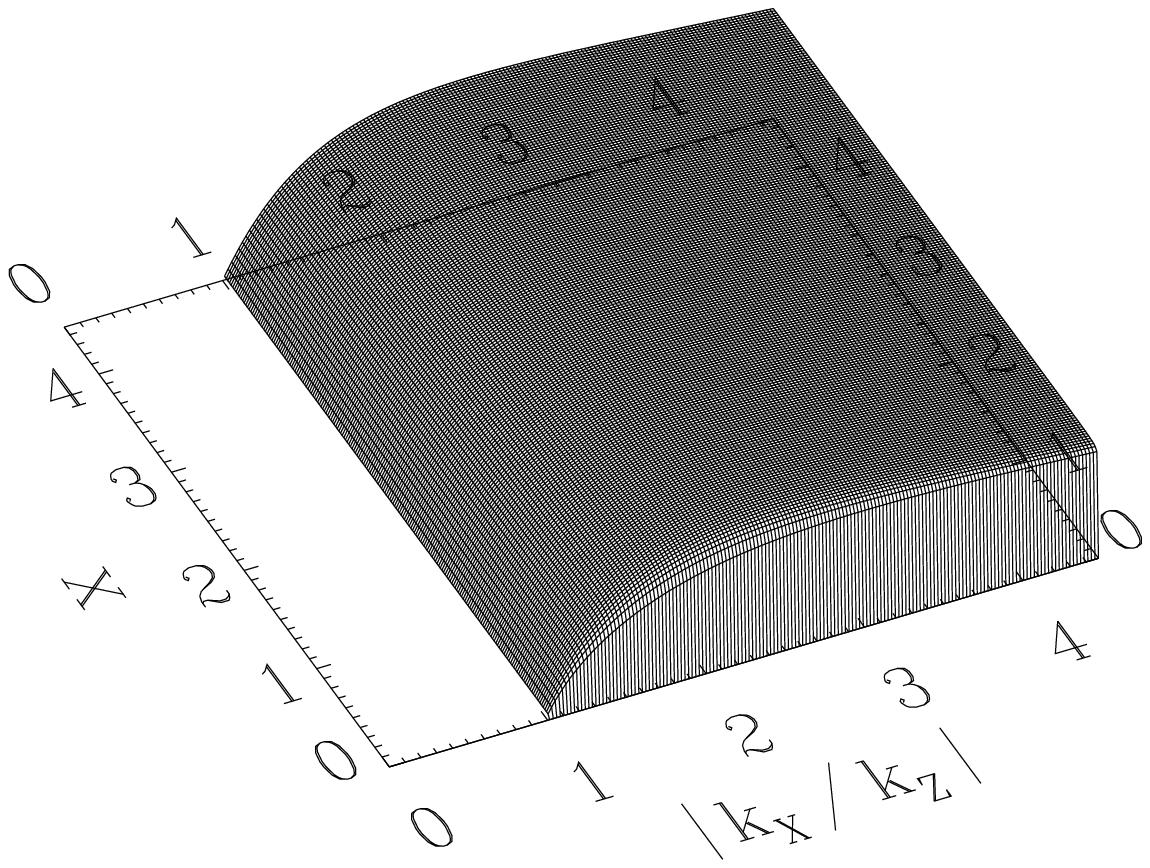}
\newline
\includegraphics[width=3.27in]{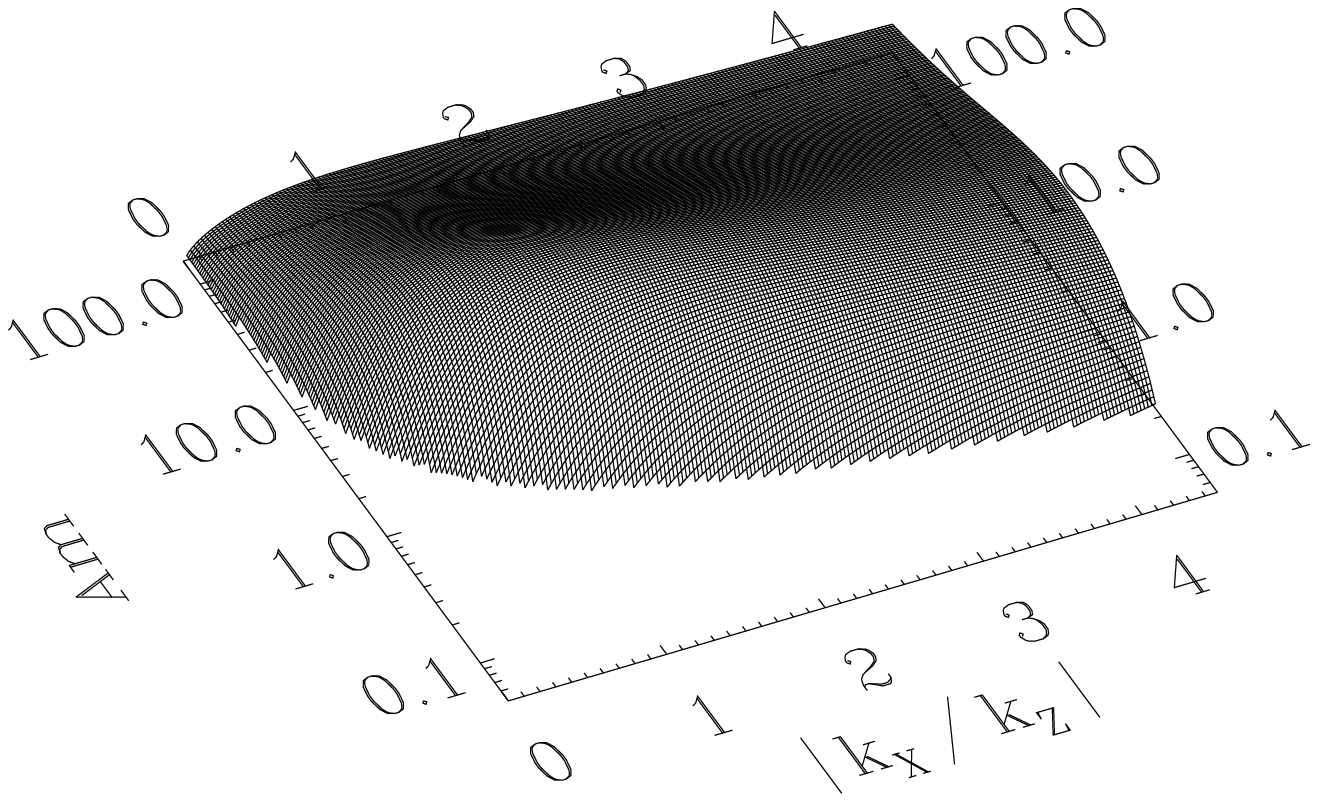}
\qquad
\includegraphics[width=3.27in]{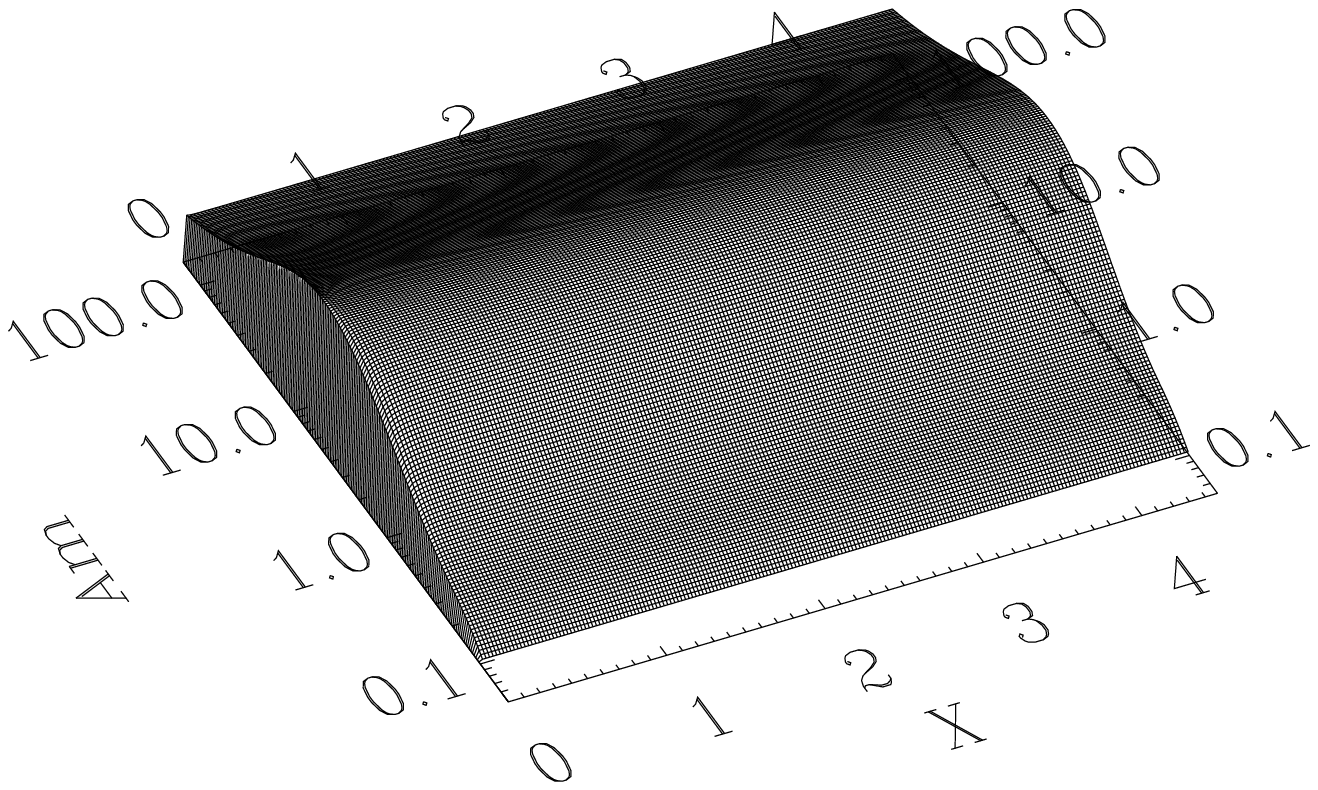}
\end{center}
\caption{{\it Counter-clockwise from top:}
ambipolar-diffusion--shear instability growth rates in (a) the
$X$-$|k_x/k_z|$ plane with ${\rm Am}=1$ and (b) the ${\rm
Am}$-$|k_x/k_z|$ plane with $X=1$; (c) maximum growth rates in the
${\rm Am}$-$X$ plane. All figures have $B_y/B_z=1$. Only regions
of instability are shown, with the height being proportional to
the growth rate. The maximum growth rate is $\sim 0.1\,|2A|$ (see
Equation \ref{E_maxgrowth}). Note that the boundary separating
stability from instability is given by Equation (\ref{E_crit2}).}
\label{F_ambigrowth}
\end{figure}

Defining the dimensionless parameters,
\begin{equation}
X \equiv \frac{\kdotva}{4A^2}\quad\quad{\rm and}\quad\quad {\rm
Am} \equiv \frac{|2A|}{\gamma\rhoi}\,,
\end{equation}
the dispersion relation may be written in dimensionless form and
growth rates may be determined numerically. In Fig.
\ref{F_ambigrowth}, we give three-dimensional plots of growth rate
in the $X$-$|k_x/k_z|$ plane (for ${\rm Am}=1$) and in the ${\rm
Am}$-$|k_x/k_z|$ plane (for $X=1$). The signs of $k_x/k_z$ and
$2A$ are chosen such that instability is possible, and $B_y/B_z$
is taken to be unity. Increasing $B_y/B_z$ does not significantly
affect the growth rates, but rather opens the available unstable
space to smaller values of ${\rm Am}$. Note that there is less
unstable parameter space as one goes to small ${\rm Am}$ (the
fluid becomes well-coupled to the magnetic field). The boundary
separating stability from instability is given by Equation
(\ref{E_crit2}). The maximum growth rate ($\sim 0.1\,|2A|$; see
Equation \ref{E_maxgrowth} below) is shown in Fig.
\ref{F_ambigrowth}c for the parameter space spanned by $X$ and
${\rm Am}$.

One final comment is worth mentioning concerning Equation
(\ref{E_crit2}). In a time-dependent zero-order state ($k_y,B_x\ne
0$), the denominator of the right-hand side grows quadratic in
time and the instability criterion will quickly become trivial to
satisfy. It is therefore of interest to rigorously test whether
this situation is realizable by performing an analysis of this
more general case.

\subsection{The General Case}

We now investigate the general case of mixed wavenumber and field
geometry via two approaches. First, we employ a WKB technique.
While this approach is strictly applicable only under special
circumstances, it will aid in the interpretation of results from
our second approach: a direct numerical solution of Equations
(\ref{E_1}) and (\ref{E_2}).

\subsubsection{WKB Analysis}\label{S_wkb}

Without loss of generality, one can express each perturbation
variable in a WKB form:
\begin{equation}\label{E_wkb}
\delta(t) \equiv \sum_{n=0}^\infty
\delta^{(n)}(t)\exp\left(i\int\omega(t)dt\right)\,,
\end{equation}
where the WKB phase has been expressed in terms of an integral
over a slowly varying frequency. This is valid as long as
$\delta^{(n+1)} \ll \delta^{(n)}$ and $d\ln\omega/dt\ll\omega$
(the adiabatic approximation). Both of these conditions are
guaranteed if $k_y/k_z \ll 1$, and we may immediately identify the
WKB parameter $k_z/k_y$. Physically, the WKB parameter represents
the ratio of the time for $k$ to change significantly to the
shearing time-scale.\footnote{There is actually an additional
condition we must impose. The resistivity tensor $\bb{\eta}$ is
dependent upon $\kxva$, and thus has an intrinsic time-dependence.
We therefore require that $\kxva$ is a slowly-varying function of
time, or equivalently, that $B_y/B_x \gg 1$. Such a large $B_y$
trivially satisfies the instability criterion.} Expressing $\delta
B_x$ and $\delta B_y$ in the form (\ref{E_wkb}), taking the limit
$k_y/k_z\rightarrow 0$, and defining $\sigma = i\omega$, the
lowest-order terms from Equations (\ref{E_1}) and (\ref{E_2})
result in the following dispersion relation\footnote{Retaining
terms of order $k_y/k_z$ contributes an extra term, similar to the
final term in equation (2.25) of \citet{bh92b}, that does not
affect the essentially oscillatory or exponential behaviour of the
solution; it is $-4Ak_xk_yk^{-2}[\sigma\tilde{\sigma}^2 +
k^2\eta_{xx}\kdotva]$, with $\tilde{\sigma}^2$ given by Equation
(\ref{E_tildesigma}). Its importance in a WKB treatment is as an
amplitude modifier on longer time scales. In a stable system, it
represents the competition between shear, which is trying to
stretch the perturbations to result in linear-in-time growth, and
ambipolar diffusion, which is trying to dampen this growth.}:
\begin{equation}\label{E_disprel2}
\sigma^4 + k^2\,{\rm tr}(\bb{\eta})\sigma^3 + \mc{C}_2\sigma^2 +
k^2\,{\rm tr}(\bb{\eta})\,\kdotva\sigma + \mc{C}_0 = 0\,,
\end{equation}
where
\renewcommand{\theequation}{\arabic{equation}\alph{subequation}}
\setcounter{subequation}{1}
\begin{equation}
\mc{C}_2 = 2\kdotva + k^4\det(\bb{\eta}) + 2Ak^2\eta_{xy}\,,
\end{equation}
\addtocounter{equation}{-1}\addtocounter{subequation}{1}
\begin{equation}
\mc{C}_0 = \kdotva\bigl[\kdotva + 2Ak^2\eta_{xy}\bigr]\,.
\end{equation}
\renewcommand{\theequation}{\arabic{equation}}
Here, ${\rm tr}(\bb{\eta})$ and $\det(\bb{\eta})$ denote the trace
and determinant, respectively, of the resistivity tensor
$\bb{\eta}$. This is similar to the dispersion relation
(\ref{E_disprel}) in \S \ref{S_case1}, and may be written in a
form similar to that of Equation (\ref{E_wavedamping}):
\renewcommand{\theequation}{\arabic{equation}\alph{subequation}}
\setcounter{subequation}{1}
\begin{equation}\label{E_wavedamping2}
\left[\sigma^2 + \sigma\Bigl(\;...\;\Bigr)_+ +
\kdotva\right]\left[\sigma^2 + \sigma\Bigl(\;...\;\Bigr)_- +
\kdotva\right] = - 2Ak^2\eta_{xy}\bigl[\sigma^2+\kdotva\bigr]\,,
\end{equation}
where \addtocounter{equation}{-1}\addtocounter{subequation}{1}
\begin{equation}
\Bigl(\;...\;\Bigr)_\pm = \frac{k^2{\rm tr}(\bb{\eta})}{2} \pm
\left[\frac{k^4{\rm tr}^2(\bb{\eta})}{4} -
k^4\det(\bb{\eta})\right]^{1/2}\,.
\end{equation}
\renewcommand{\theequation}{\arabic{equation}}
Evidently, the effect of shear is to couple different
polarizations of Alfv\'{e}n waves, which are damped at different
rates. It is also of interest to calculate the associated
eigenvectors. Defining
\begin{equation}\label{E_tildesigma}
\tilde{\sigma}^2 \equiv \sigma^2 + k^2{\rm tr}(\bb{\eta})\sigma +
k^4\det(\bb{\eta}) + \kdotva + 2Ak^2\eta_{xy}\,,
\end{equation}
the (non-trivial) eigenvector components (in the limit $k_y/k_z
\ll 1$) can be expressed as
\renewcommand{\theequation}{\arabic{equation}\alph{subequation}}
\setcounter{subequation}{1}
\begin{equation}\label{E_eigenvy}
\delta v_y = \frac{\delta v_x}{\sigma\tilde{\sigma}^2 +
k^2\eta_{xx}\kdotva}\,\bigl[-2A\tilde{\sigma}^2 + \kdotva
k^2\eta_{yx}\bigr]\,,
\end{equation}
\addtocounter{equation}{-1}\addtocounter{subequation}{1}
\begin{equation}\label{E_eigenby}
\delta B_{y} = \frac{-i(\bb{k}\bcdot\bb{B})\,\delta
v_x}{\sigma\tilde{\sigma}^2 + k^2\eta_{xx}\kdotva}\, \bigl[
k^2\eta_{yx}\sigma + 2Ak^2\eta_{xx}\bigr]\,,
\end{equation}
\addtocounter{equation}{-1}\addtocounter{subequation}{1}
\begin{equation}\label{E_eigenbx}
\delta B_{x} = \frac{i(\bb{k}\bcdot\bb{B})\,\delta
v_x}{\sigma\tilde{\sigma}^2 + k^2\eta_{xx}\kdotva}\,
\bigl[\sigma^2 + k^2\eta_{yy}\sigma + \kdotva +
2Ak^2\eta_{xy}\bigr] \,.
\end{equation}
\renewcommand{\theequation}{\arabic{equation}}
As in \S \ref{S_case1}, a sufficient condition for instability is
$\mc{C}_0<0$, or
\begin{equation}\label{E_crit3}
\kdotva + 2Ak^2\eta_{xy} < 0\,.
\end{equation}
This criterion depends on both $k_x(t)$ and $B_y(t)$, through
$\eta_{xy}$. As a result, the destabilizing term will grow in
amplitude, opening up more and more unstable parameter space as
time progresses and increasing the growth rate.

If we view the dispersion relation (\ref{E_disprel2}) as an
equation in $k^2$ and $\kdotva$, the maximum growth rate can be
calculated for a given $\bb{\eta}$. At the maximum growth rate
$\sigma = \sigma_{\rm max}$, partial differentiation of Equation
(\ref{E_disprel2}) with respect to $k^2$ and $\kdotva$ gives the
two equations
\begin{equation}\label{E_partial1}
k^2{\rm tr}(\bb{\eta})\sigma^3_{\rm max} +
\bigl[2k^4\det(\bb{\eta}) + 2Ak^2\eta_{xy}\bigr]\sigma^2_{\rm max}
+ k^2{\rm tr}(\bb{\eta})\,\kdotva\sigma_{\rm max} +
2Ak^2\eta_{xy}\,\kdotva = 0\,,
\end{equation}
\begin{equation}\label{E_partial2}
2\sigma^2_{\rm max} + k^2{\rm tr}(\bb{\eta})\sigma_{\rm max} +
2\kdotva + 2Ak^2\eta_{xy} = 0\,.
\end{equation}
Eliminating $\kdotva$ between these two leads after regrouping to
a surprisingly simple result,
\begin{equation}
\bigl[4\det(\bb{\eta}) - {\rm tr}^2(\bb{\eta})\bigr]\sigma^2_{\rm
max} - 4A\eta_{xy}{\rm tr}(\bb{\eta})\sigma_{\rm max} -
4A^2\eta^2_{xy} = 0\,.
\end{equation}
There are two solutions to this equation, corresponding to the
extrema of the dispersion relation. Only one of these is a
physically meaningful solution satisfying the dispersion relation
(\ref{E_disprel2}); it is given by
\begin{equation}\label{E_maxgrowth}
\sigma_{\rm max} =
|A|\,\left|\frac{\eta_{xy}}{\det^{1/2}(\bb{\eta})
+\frac{1}{2}\,{\rm tr}(\bb{\eta})}\right|\,.
\end{equation}
This is a remarkable result. \citet{bh92a} conjectured that the
maximum growth rate of any instability feeding off the
differential rotation in a disc is given by the local Oort $A$
value, $\sigma_{\rm A}\equiv (1/2)|d\Omega/d\ln R|$, no matter the
cause for instability. The reason is rooted in the dynamics of the
differential rotation itself. In this paper, we are concerned with
{\em planar} shear flows, and we arrive at a similar result, with
the shear playing the role of the differential rotation. It is
notable that Equation (\ref{E_maxgrowth}) is independent of the
degree of ionization, depending only upon the geometry of the
background magnetic field and the shearing rate $|2A|$.

The next order in a WKB expansion yields the time dependence of
the slowly-varying amplitude. Provided we are able to compute this
amplitude, the eigenvectors may be used to compute the wave energy
and compare with the numerical results given in the next section.
This has been done, e.g., by Johnson (2007) for the case of
nonaxisymmetric shearing waves in differently-rotating disks. In a
non-dissipative, continuous system, the amplitude may be computed
from conservation of wave action. In the presence of ambipolar
diffusion, however, wave action is not conserved, and retrieving
the amplitude in this fashion is prohibitive. Instead, one must
take the algebraically tedious approach of directly calculating
the higher-order expansion for the modes. We have done this,
finding an equation similar to that of equation (A5) of
\citet{johnson07}. Unfortunately, once the eigenvectors
(\ref{E_eigenvy}) - (\ref{E_eigenbx}) are substituted in, the
result cannot be easily integrated and the slowly-varying
amplitude cannot be calculated. Numerical solutions seem to be the
most profitable approach.

\subsubsection{Numerical Solution}\label{S_numerical}

Here we undertake a direct numerical solution of Equations
(\ref{E_1}) and (\ref{E_2}). Following lines similar to those
developed in \citet{glb65}, we introduce a new independent time
variable
\begin{equation}
\tau \equiv \frac{k_x(t)}{k_z} = \frac{k'_x}{k_z} -
2A\frac{k_y}{k_z}\,t\,,
\end{equation}
so that $(k/k_z)^2 = \tau^2 + (k_y/k_z)^2 + 1$. Equations
(\ref{E_1}) and (\ref{E_2}) may then be written in dimensionless
form and numerically integrated. All that remains is to specify
initial values for $\delta B_x$, $\delta B_y$, $\delta B_z$, and
$B_y/B_z$. (The initial value of $\tau$ is determined from
$\bb{k}\bcdot\delta\bb{B}=0$.) Unfortunately, we must also specify
values for $k_y/k_z$ and $B_x/B_z$. However, we have undertaken a
parameter study to see if varying these significantly influences
the results, and have found that the qualitative behaviour is not
affected. The results presented here have $\delta B_x(0) = 0.3$,
$\delta B_y(0) = \delta B_z(0) = 0$, and
$k_y/k_z=B_x/B_z=B_y(0)/B_z=1$.

\begin{figure}
\includegraphics[width=3.35in]{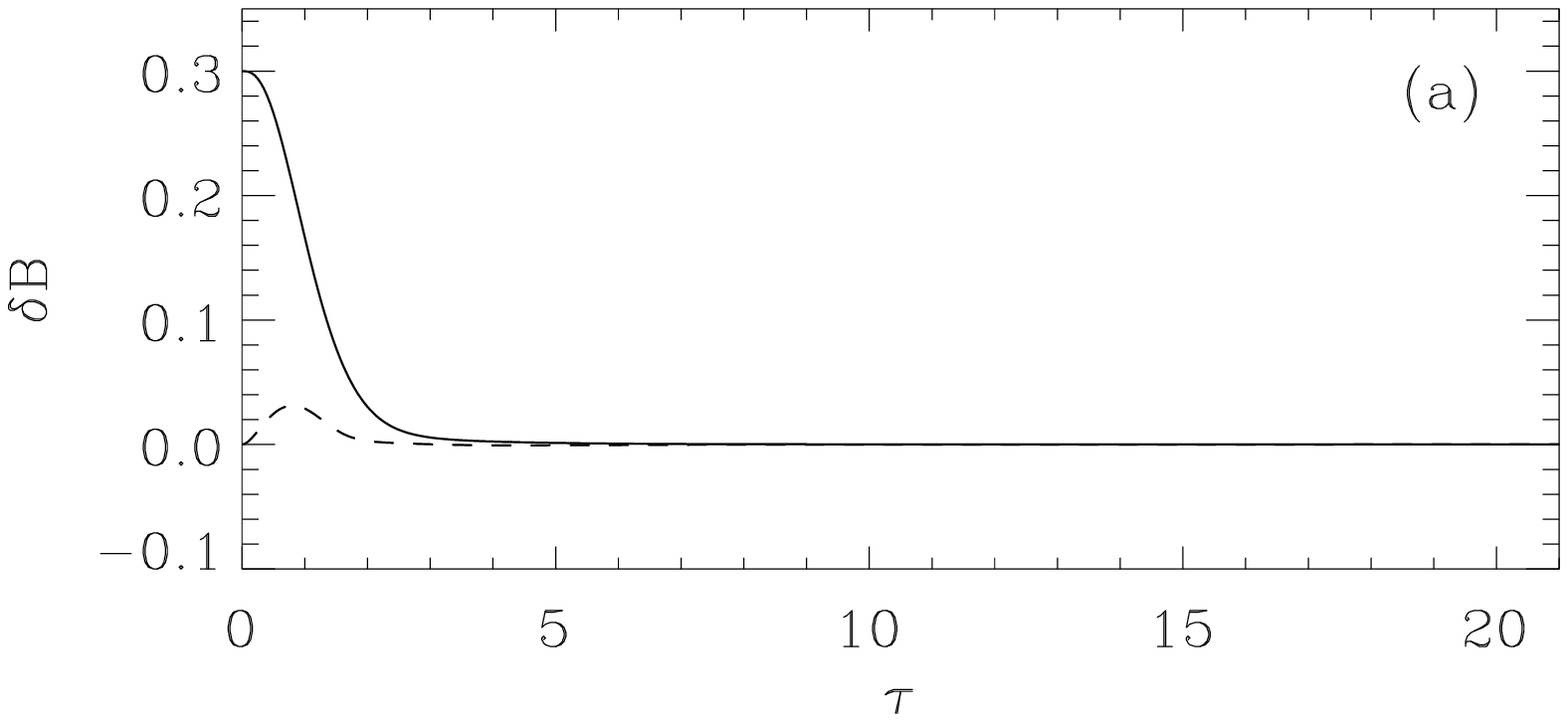}
\quad\quad
\includegraphics[width=3.35in]{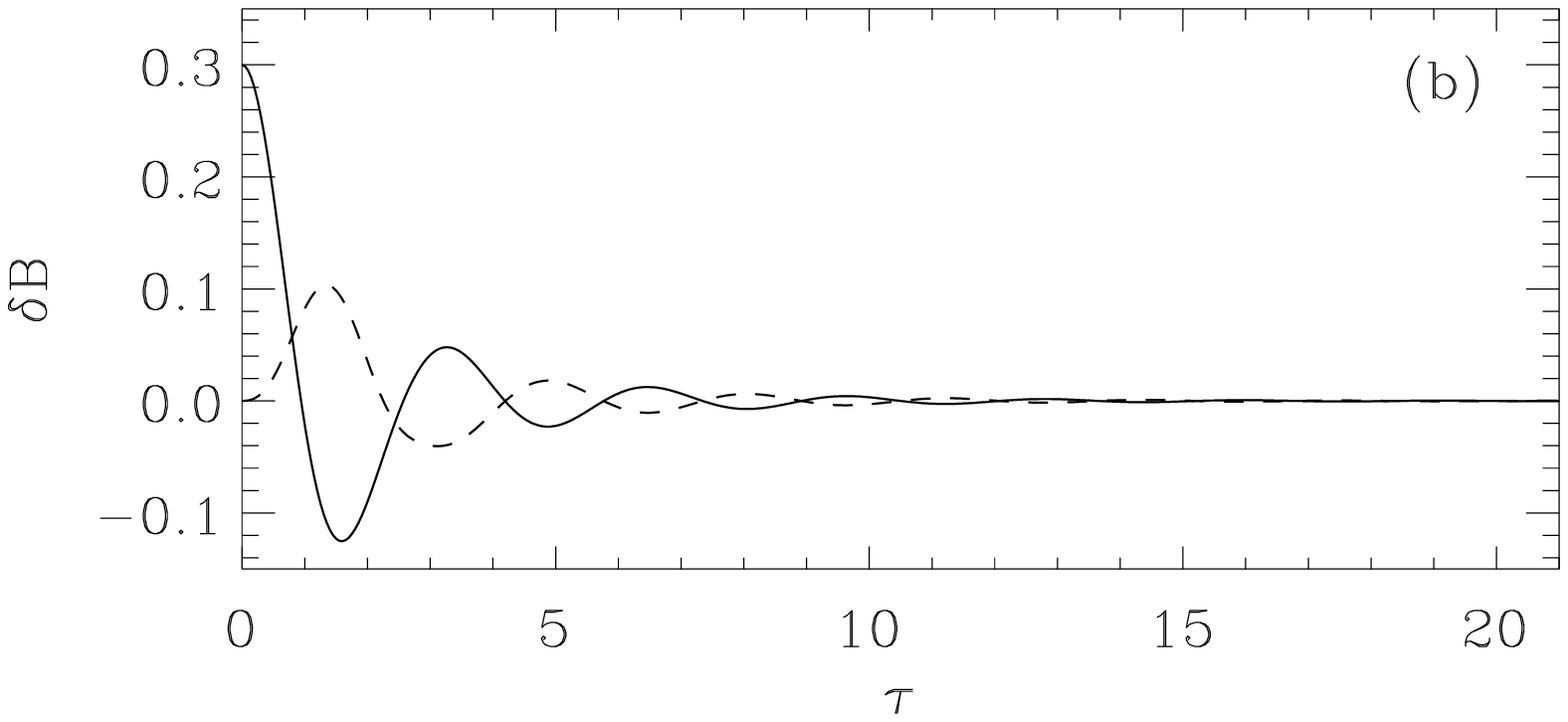}
\newline
\newline
\includegraphics[width=3.35in]{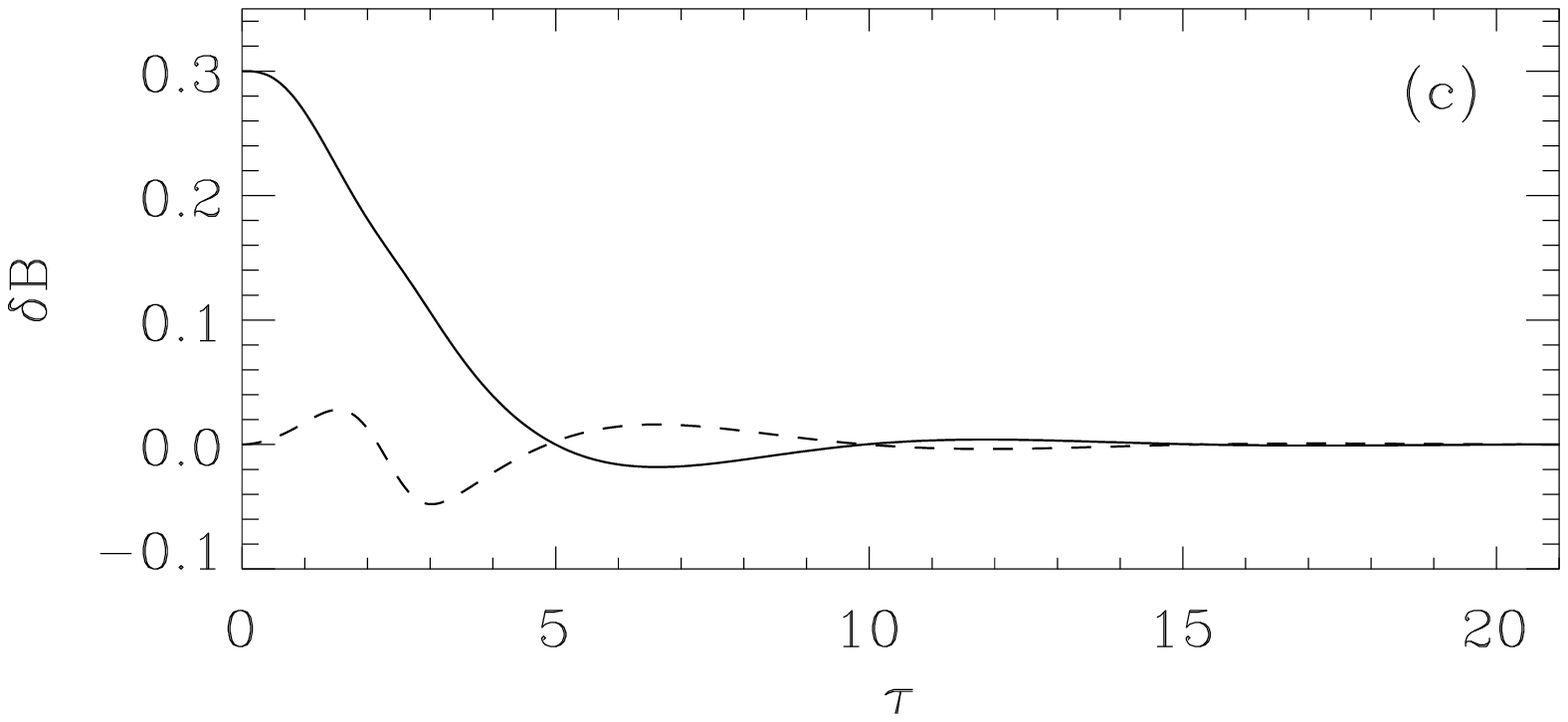}
\quad\quad
\includegraphics[width=3.35in]{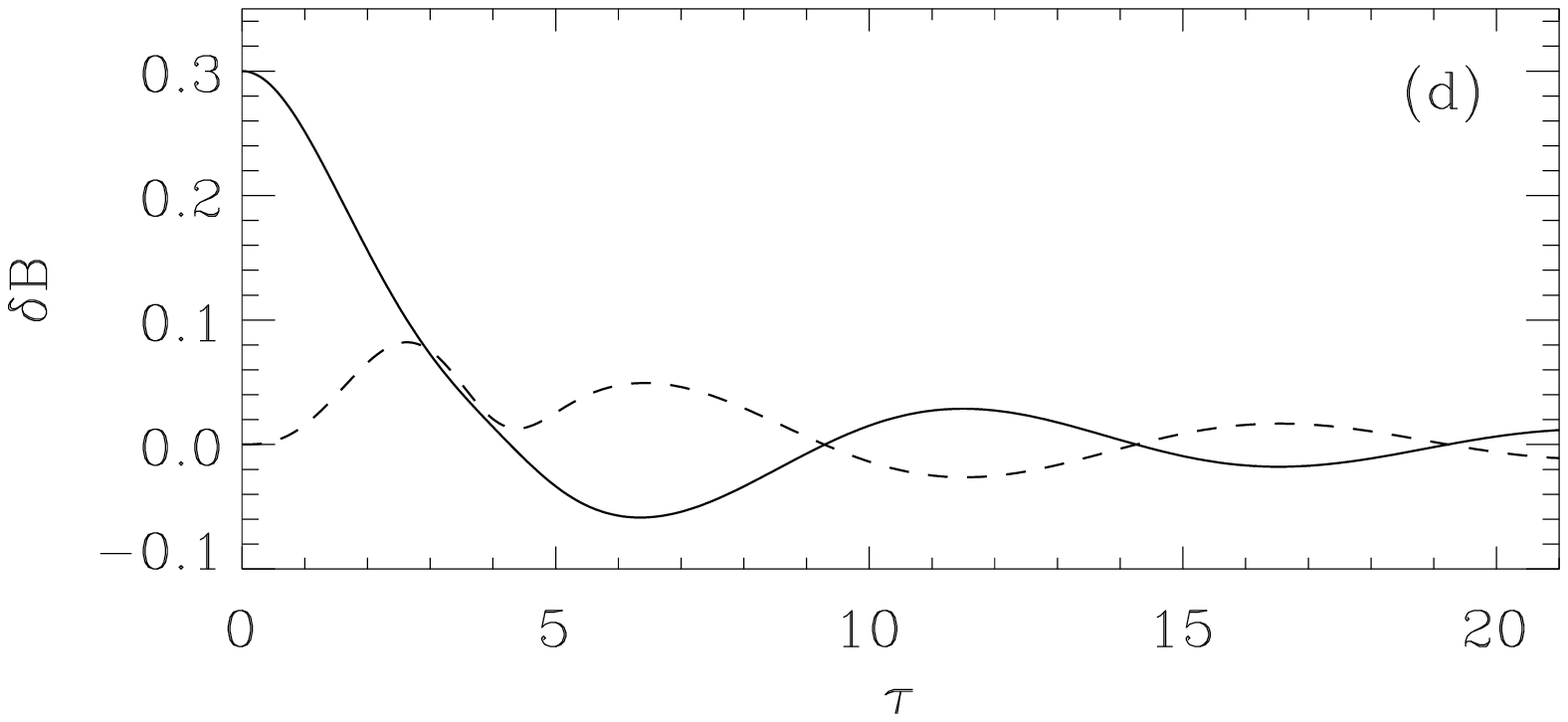}
\caption{Evolution of the perturbed magnetic fields $\delta B_x$
(solid line) and $\delta B_y$ (dashed line) for (a) $(X,\,{\rm
Am})=(1.0,\,1.0)$ , (b) $(X,\,{\rm Am})=(1.0,\,0.1)$, (c)
$(X,\,{\rm Am})=(0.1,\,1.0)$, and (d) $(X,\,{\rm
Am})=(0.1,\,0.1)$. The sign of $2Ak'_x/k_z$ is chosen to be
negative so that the flow is stable. The initial amplitudes are
$\delta B_x=0.3$ and $\delta B_y=\delta B_z=0$, and $k_y/k_z =
B_x/B_z = B_y(0)/B_z = 1$. The initial value of $\tau=k'_x/k_z$ is
determined by $\bb{k}\bcdot\delta\bb{B}=0$.} \label{F_bxst}
\end{figure}

In Fig. \ref{F_bxst}, we give the evolution of $\delta B_x$ (solid
line) and $\delta B_y$ (dashed line) for (a) $(X,\,{\rm
Am})=(1.0,\,1.0)$ , (b) $(X,\,{\rm Am})=(1.0,\,0.1)$, (c)
$(X,\,{\rm Am})=(0.1,\,1.0)$, and (d) $(X,\,{\rm
Am})=(0.1,\,0.1)$. Here, the sign of $2Ak'_x/k_z$ is chosen to be
negative so that the instability does not operate. The modes are
damped due to ambipolar diffusion at a rate proportional to
$2A/\gamma\rhoi$ (i.e., a longer neutral-ion collision time-scale
results in faster ambipolar diffusion). Decreasing $X$ results in
a smaller damping rate (i.e., a longer Alfv\'{e}n crossing
time-scale allows more time for an ion to communicate the presence
of the magnetic field to a neutral via collisions).

\begin{figure}
\includegraphics[width=3.35in]{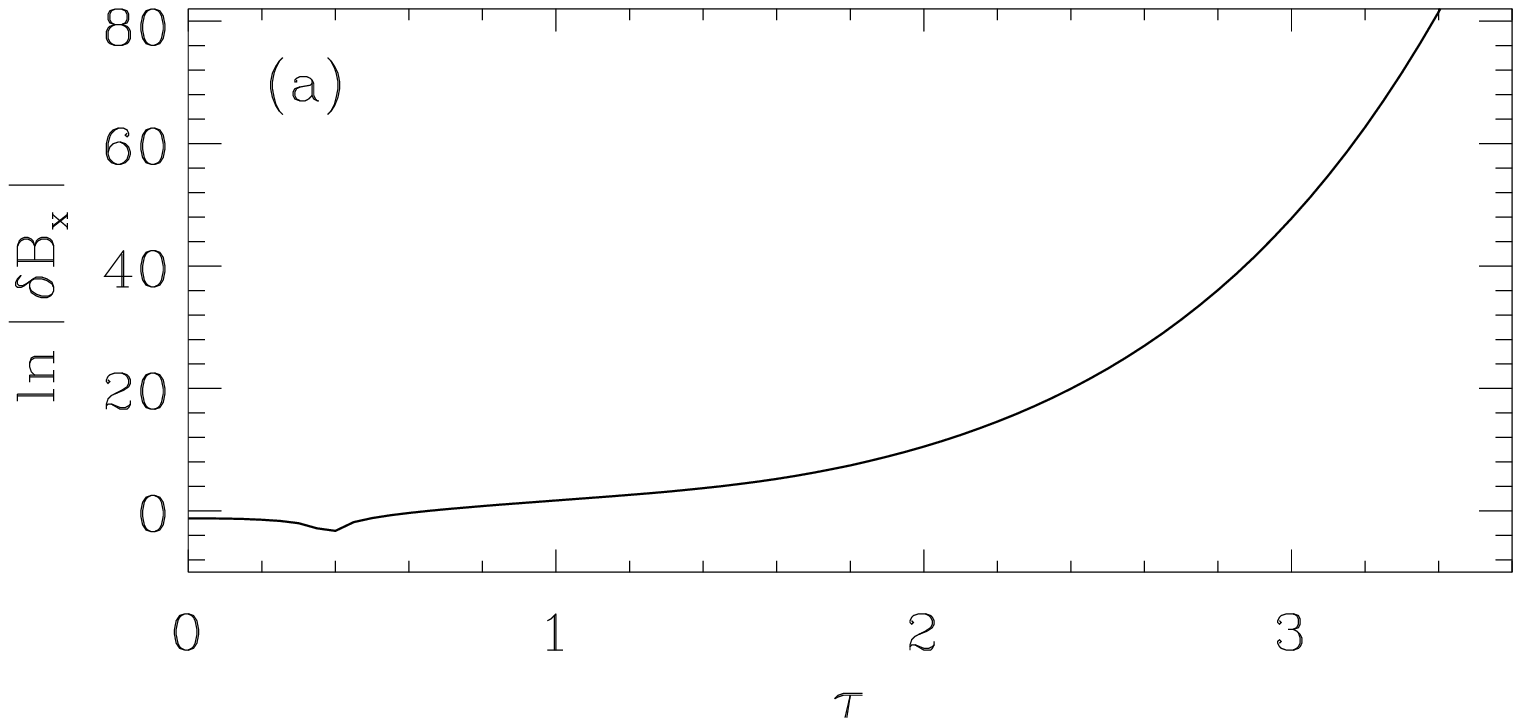}
\quad\quad
\includegraphics[width=3.35in]{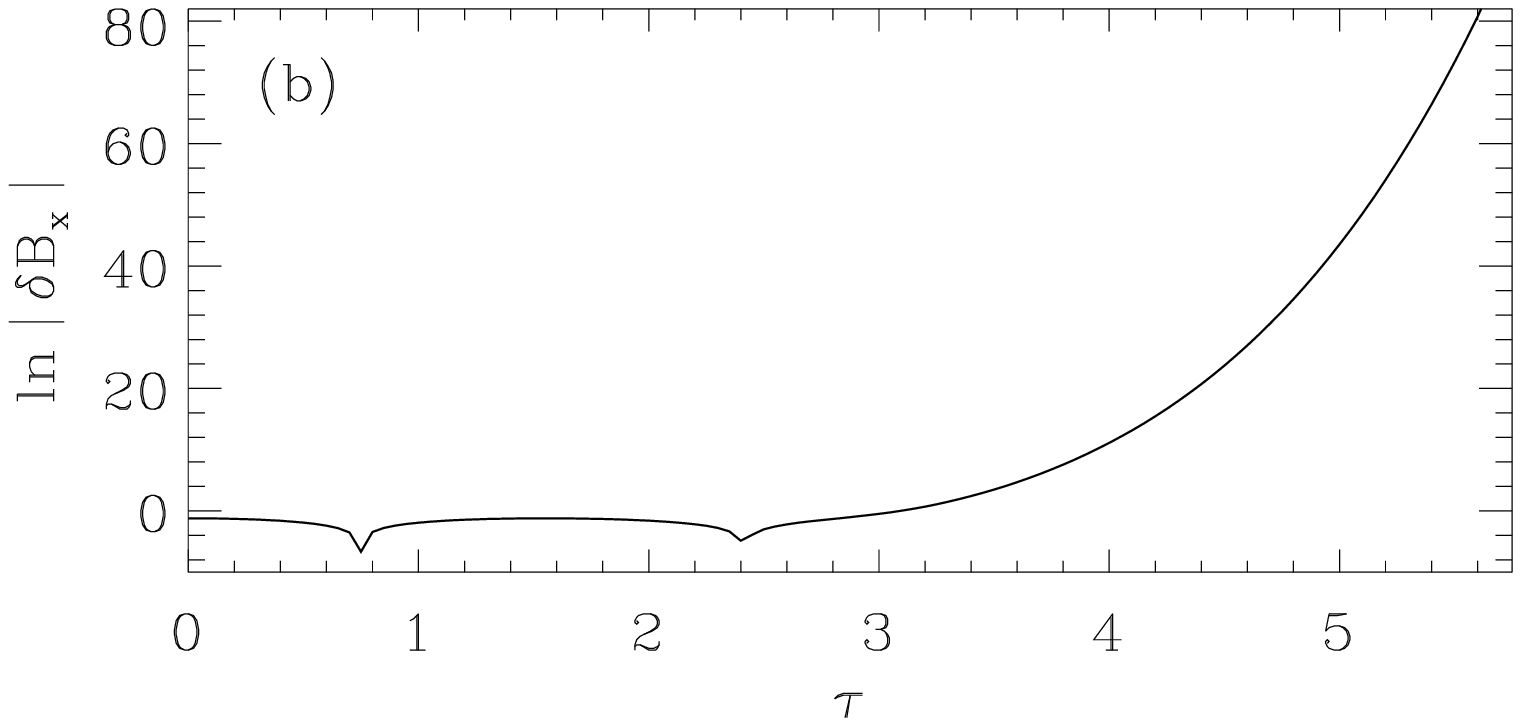}
\newline
\newline
\includegraphics[width=3.35in]{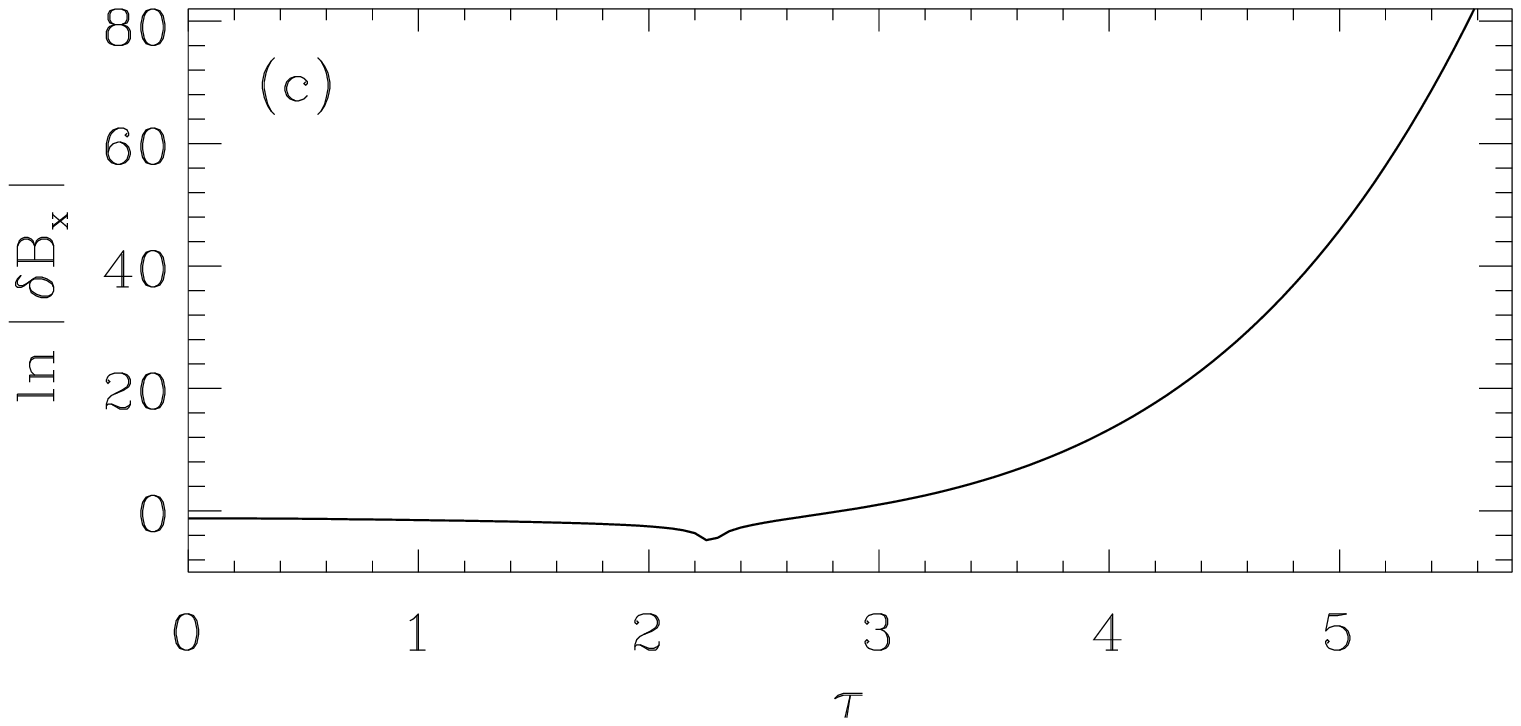}
\quad\quad
\includegraphics[width=3.35in]{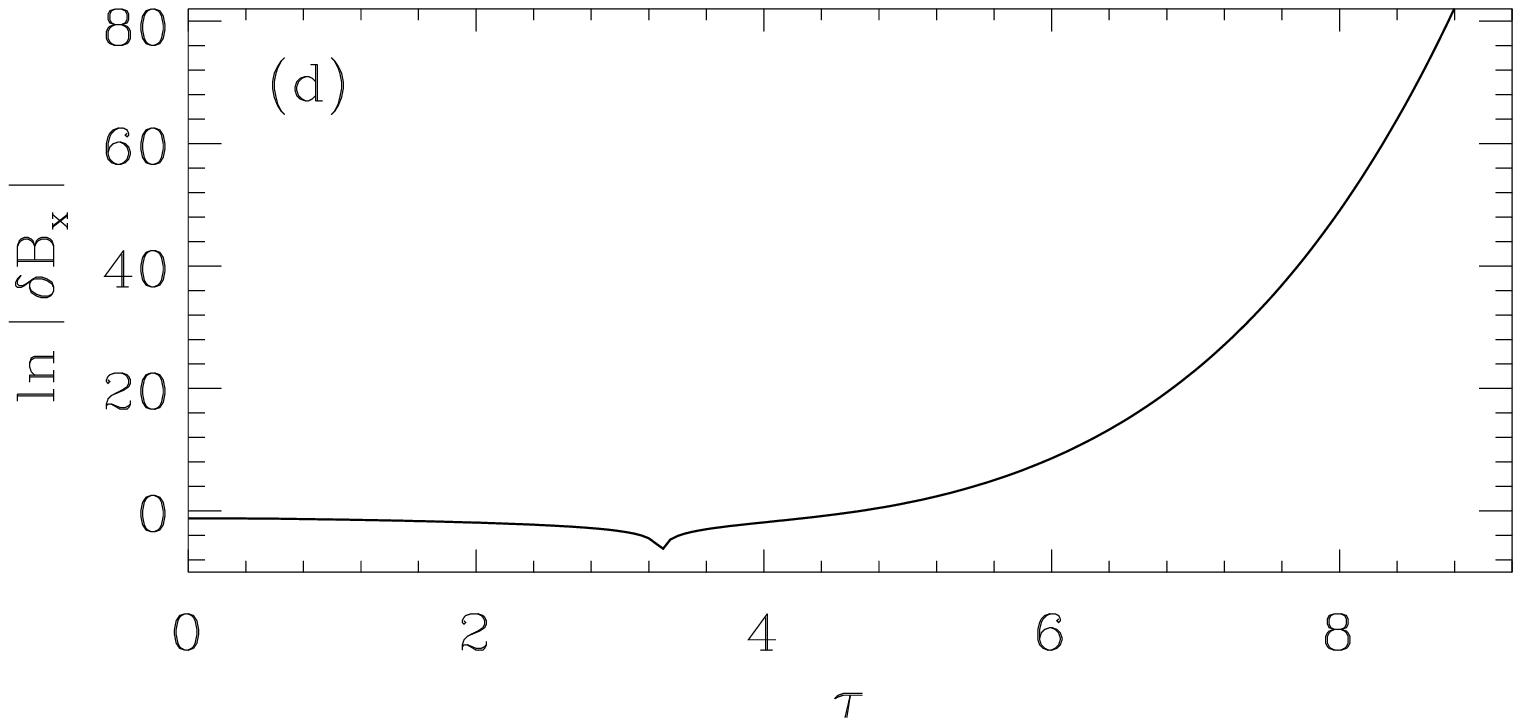}
\caption{Same as in Fig. \ref{F_bxst}, except that the sign of
$2Ak'_x/k_z$ is chosen to be positive so that the flow is
unstable. Note that the ordinate value is $\ln|\delta B_x|$,
rather than $\delta B_x$ as in Fig. \ref{F_bxst}. The evolution of
$\delta B_y$ is similar and therefore not shown. There are, in
general, two stages of evolution. Before the instability criterion
is satisfied (i.e., $\tau=k_x(t)/k_z$ is too small) we have wave
damping due to ambipolar diffusion. Once $\tau=k_x(t)/k_z$ reaches
the value where Equation (\ref{E_crit3}) is marginally satisfied,
the growth phase begins.}\label{F_bxun}
\end{figure}

In Fig. \ref{F_bxun}, we give the evolution of the corresponding
unstable modes for the same parameters as in Fig. \ref{F_bxst}.
The evolution of $\delta B_y$ in this case is similar to $\delta
B_x$, and therefore is not given. Note that $\ln|\delta B_x|$ is
plotted, rather than $\delta B_x$, so that the growth may be more
clearly seen. There are, in general, two stages of evolution.
Before the instability criterion is satisfied (i.e.,
$\tau=k_x(t)/k_z$ is too small) we have wave damping due to
ambipolar diffusion (i.e., $\ln|\delta B_x|<0$). (The cusps on the
plots correspond to the wave amplitude passing through zero and
changing sign. For example, in Fig. \ref{F_bxun}b, the wave
executes one full cycle before becoming unstable.) Once
$\tau=k_x(t)/k_z$ reaches the value where Equation (\ref{E_crit3})
is marginally satisfied, the growth phase begins (i.e.,
$\ln|\delta B_x|>0$).

\begin{figure}
\includegraphics[width=3.35in]{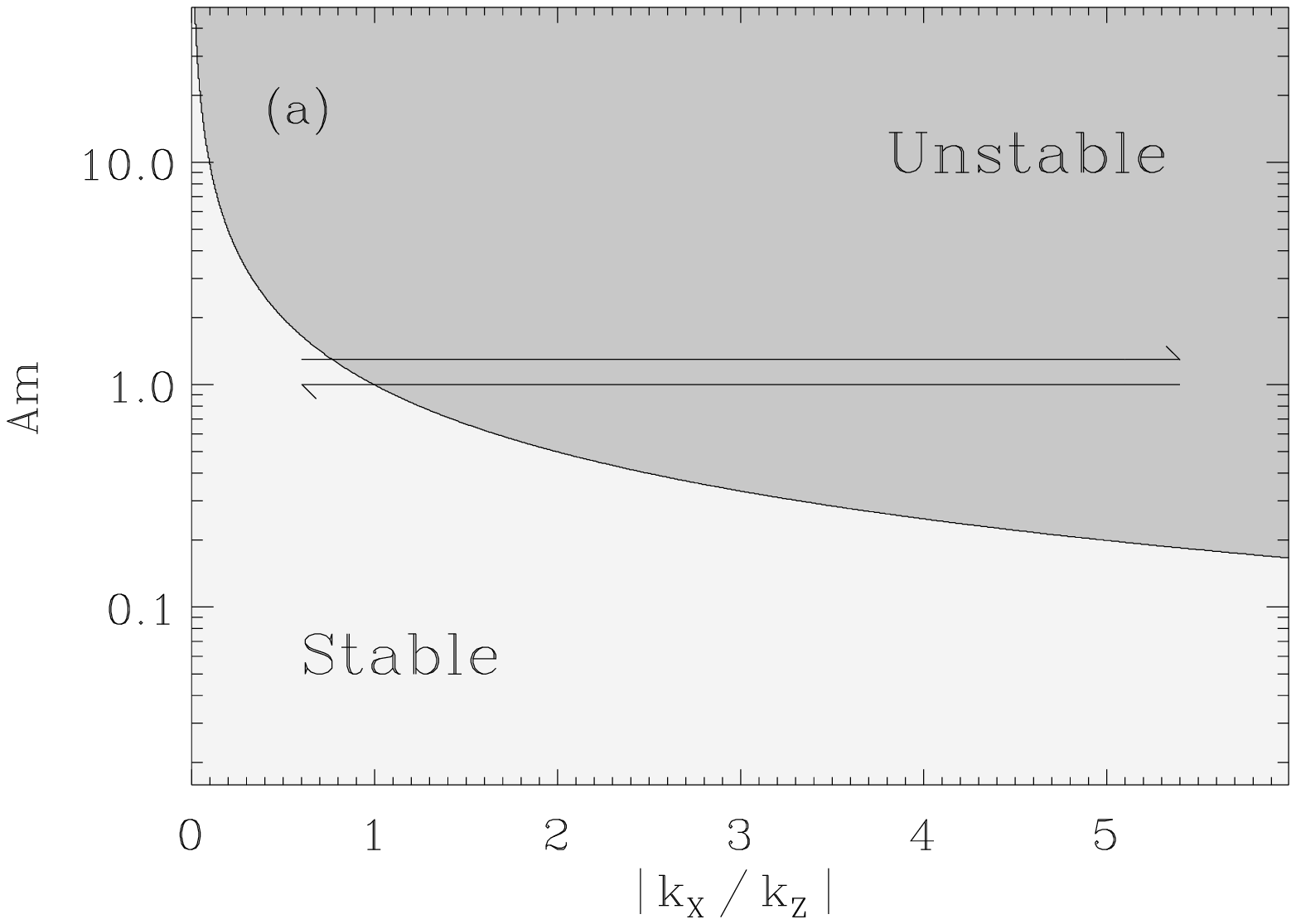}
\quad\quad
\includegraphics[width=3.35in]{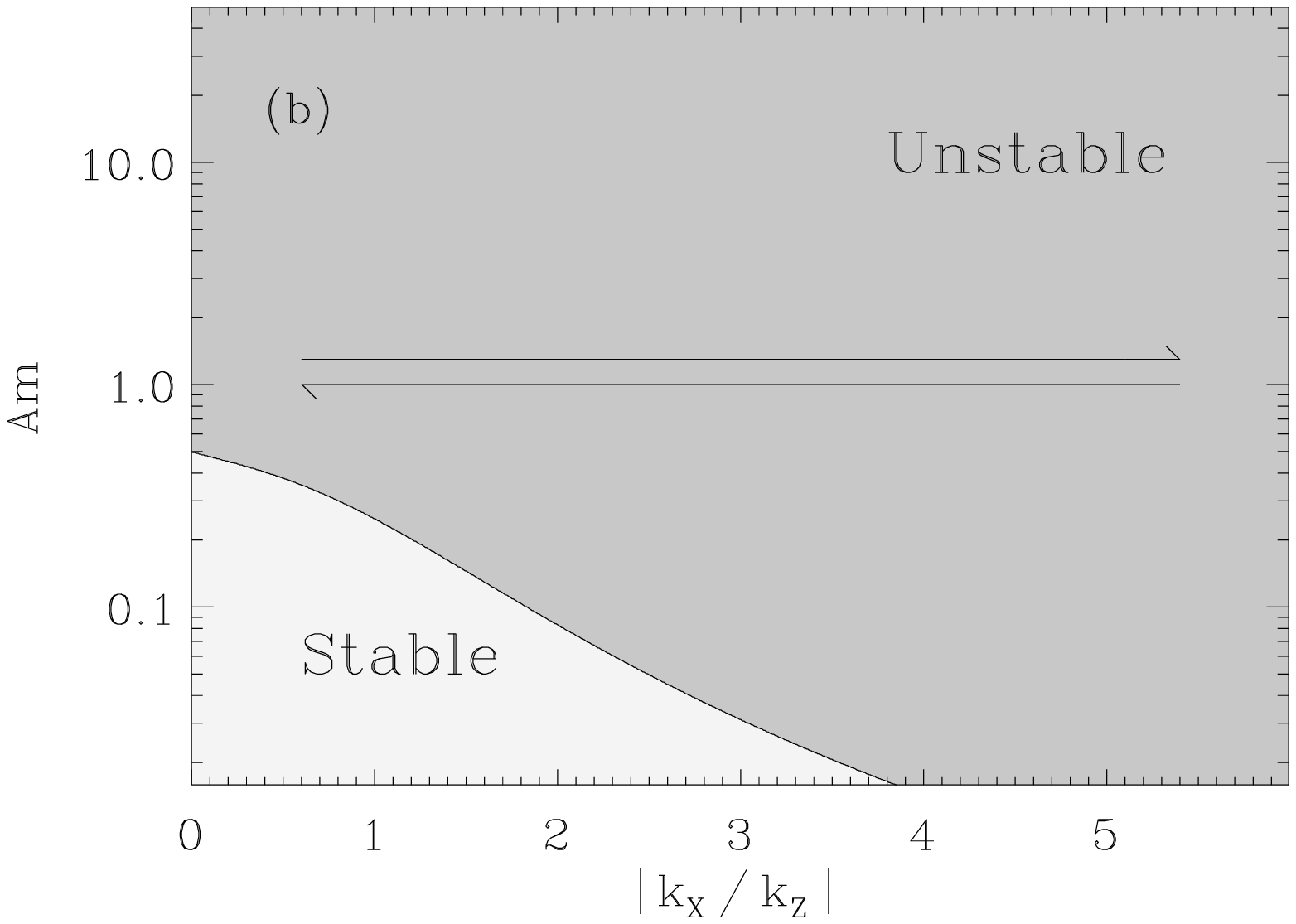}
\caption{(a) Regions of stability and instability in the ${\rm
Am}$-$|k_x/k_z|$ plane for $B_x=0$. Initially, $|k_x/k_z|$ is
large, and the point defining the wavevector moves to the left
through the unstable region on a constant ${\rm Am}$ line. If
$\kdotva > -2Ak^2\eta_{xy}$, then a finite portion of time is
spent in the stable region. After attaining its minimum (i.e.,
$k_x/k_z=0$), the wavevector point retraces its path to the right,
reentering the unstable region. (b) Same as in (a), but for
$B_x/B_z = 1$. See text for discussion.}\label{F_stability}
\end{figure}

Similar to investigations of other instabilities in the presence
of a shearing time-dependent background (e.g., Balbus \& Hawley
1992b, Balbus \& Terquem 2001), we find that the evolution unfolds
as a series of time-independent problems, whose analysis is aided
by the existence of WKB solutions. Here, however, the analysis is
complicated somewhat by the explicit time-dependence of the
unstable terms. To keep things simple at first, consider the case
$B_x=0$, so that the time-evolution of the destabilizing term is
dependent only upon the wavenumber $k_x(t) (\,= k'_x - 2Ak_y t)$.
In Fig. \ref{F_stability}a we show the evolution of an initially
leading wavenumber through the parameter space defined by ${\rm
Am}$ and $|k_x/k_z|$ for $k_y/k_z=1$ and $B_y(\tau)=B_y(0)=0$. The
wavevector evolution traces (and retraces) a horizontal line in
the plane of Fig. \ref{F_stability}a. In contrast to the
nonaxisymmetric MRI \citep{bh92b}, the wavevector evolves {\em
towards} stability, not away from it, and in principle spends only
a finite amount of time in the stable region, that is, when
$\kdotva > -2Ak^2\eta_{xy}$ is satisfied. In Fig.
\ref{F_stability}b, we relax our assumption of a vanishing $B_x$
and take $B_x/B_z=1$, so that $B_y$ is time-dependent. Recall that
$k_x/k_z$ is our time parameter, so that $B_y$ increases with
$k_x$. One modification to the picture presented in Fig.
\ref{F_stability}a is an extension of the unstable region at large
$|k_x/k_z|$. This is due to the time-dependence of $B_y$ (and
thereby, $\eta_{xy}$). Another important consequence is that, in
principle, the growth rates will increase in time due to the fact
that a strongly trailing (or leading) wavevector is trivially
unstable. (These modes are well-localized in a WKB sense.) If the
exponential growth phase goes unchecked by non-linear effects, it
is conceivable that the growth rate will quickly surpass the
shearing rate, or for that matter, any other dynamical rate in a
system of interest.

\begin{figure}
\includegraphics[width=3.35in]{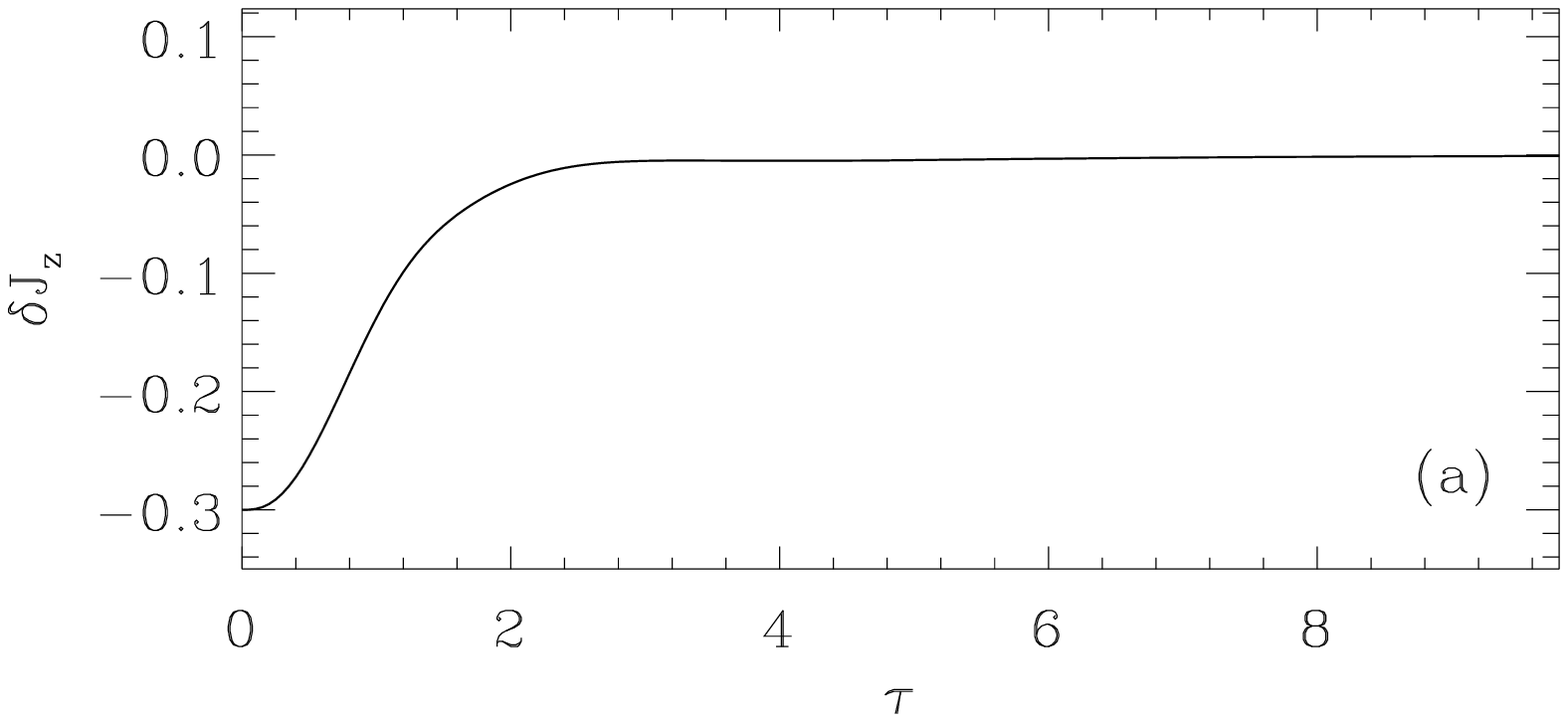}
\quad\quad
\includegraphics[width=3.35in]{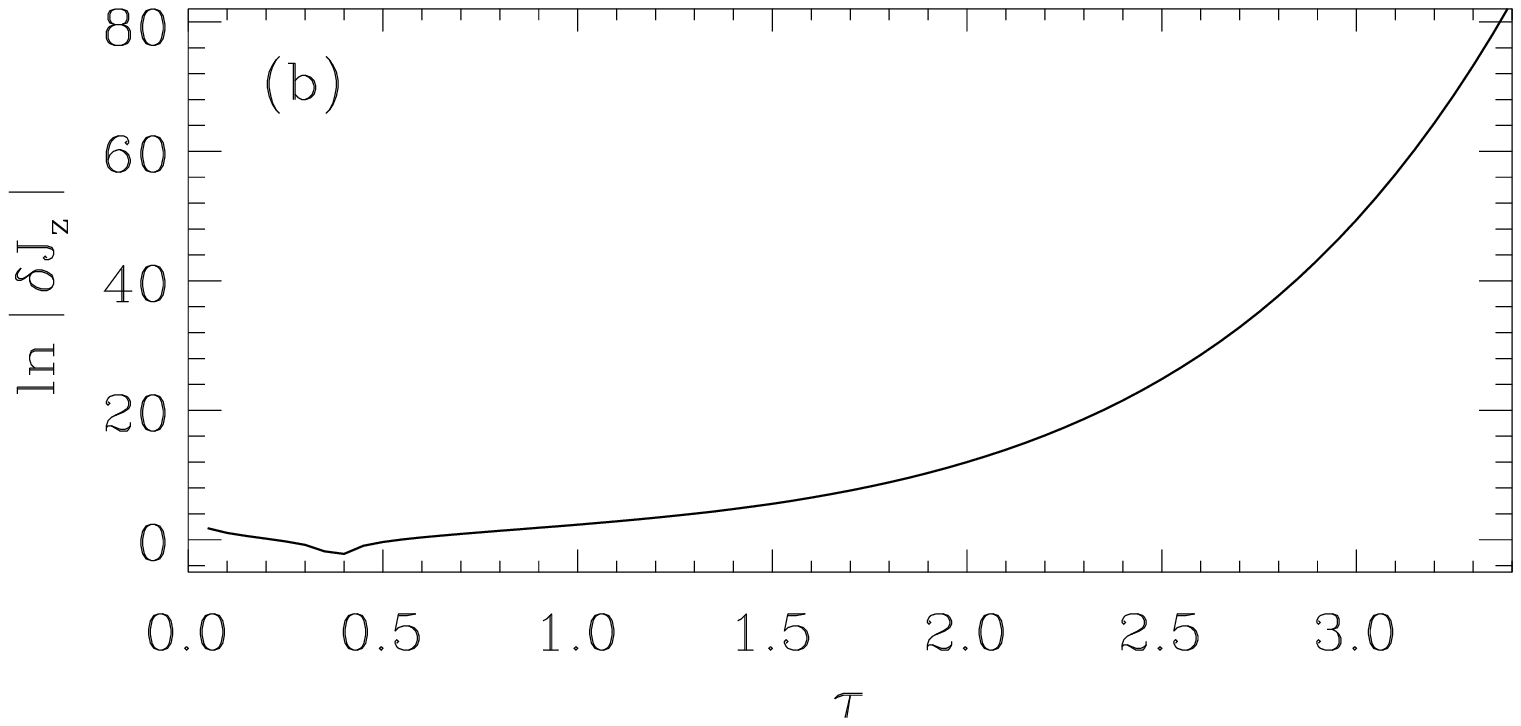}
\caption{Evolution of (a) $\delta J_z$ for
$k^2\bb{\eta}\,\bb{:}\,\del\,\bb{v} > 0$ (stable), and (b)
$\ln|\delta J_z|$ for $k^2\bb{\eta}\,\bb{:}\,\del\,\bb{v} < 0$
(unstable). Here, $(X,\,{\rm Am})=(1.0,\,1.0)$; initial values are
as in Figures \ref{F_bxst} and \ref{F_bxun}.\label{F_current}}
\end{figure}

One final issue remains: where does the energy come from to
sustain this growth? The last term in Equation (\ref{E_crit3})
introduces a novel form of coupling: in dyadic notation, it is
\[
k^2\bb{\eta}\,\bb{:}\,\del\,\bb{v}\,.
\]
This suggests that ambipolar diffusion influences the free energy
path between the velocity shear and the perturbations. In an ideal
MHD shear flow, where ambipolar diffusion is absent, the link
between the fluctuations and the free energy source is provided by
vortex stretching. Magnetic fields modify this picture, of course,
providing what amounts to an effective surface tension by
resisting the shear. The effect of ambipolar diffusion may be most
easily seen by using Equations (\ref{E_1}) and (\ref{E_2}) to
derive an equation for
\begin{equation}
\delta J_z \equiv k_x\delta B_y - k_y\delta B_x\,,
\end{equation}
which is proportional to the $z$-component of the perturbed
current density:
\begin{eqnarray}\label{E_current}
\lefteqn{\frac{1}{k_x}\left[\frac{d^2}{dt^2} +
\kdotva\right]\delta J_z + \frac{4Ak_y}{k_x}\frac{d\delta
B_y}{dt}}\nonumber\\*&&\mbox{} =
-\frac{d}{dt}\bigl(k^2\eta_{yy}\delta B_y + k^2\eta_{yx}\delta
B_x\bigr) + \frac{k_y}{k_x}\frac{d}{dt}\bigl(k^2\eta_{xx}\delta
B_x + k^2\eta_{xy}\delta B_y\bigr) - 2A\bigl(k^2\eta_{xx}\delta
B_x + k^2\eta_{xy}\delta B_y\bigr) \,.
\end{eqnarray}
In the absence of ambipolar diffusion, the right-hand side
vanishes, and we are left with a wave equation for $\delta J_z$
with the additional effect of shear acting as an amplitude
modifier. The right-hand side is best approached piecemeal. The
first two terms represent the damping of the current due to
ambipolar diffusion. It is clear that different components of the
perturbed current are damped differently. If ambipolar diffusion
acted isotropically, with a resistivity
$\eta\equiv\eta_{xx}=\eta_{yy}$, these terms would read
\[
-\frac{1}{k_x}\frac{d}{dt}\,k^2\eta\delta J_z -
\frac{2Ak_y}{k_x}\,k^2\eta\delta B_y\,,
\]
which shows the effect of resistivity on the propagation of the
wave, as well as the attempt to resist the stretching of the
perturbed magnetic field by shear. The final term in Equation
(\ref{E_current}) is the key to instability. It becomes a source
term when $k^2\bb{\eta}\,\bb{:}\,\del\,\bb{v} < 0$, extracting
energy from the background velocity shear. In Fig.
\ref{F_current}, we show the evolution of $\delta J_z$ for (a)
$k^2\bb{\eta}\,\bb{:}\,\del\,\bb{v} > 0$ (stable), and (b)
$k^2\bb{\eta}\,\bb{:}\,\del\,\bb{v} < 0$ (unstable). For the
unstable case, the $z$-component of the vorticity ($\equiv
k_x\delta v_y - k_y\delta v_x$) follows a similar evolution to the
current.

\section{Hall--Shear Instability}\label{S_hall}

We now consider the case where the Hall effect is the dominant
non-ideal MHD process, corresponding to a neutral number density
much in excess of $10^{13}$ cm$^{-3}$. The analysis follows the
same course as in \S\,\ref{S_AD}, provided we redefine the
resistivity tensor to be
\begin{eqnarray}\label{E_eta2}
k^2\bb{\eta} \rightarrow \hall \left(
\begin{array}{rr}
k_x k_y/k^2_z & k^2_{yz}/k^2_z \\
-k^2_{xz}/k^2_z & -k_xk_y/k^2_z
\end{array}
\right)\,.
\end{eqnarray}
Realizing that $k^2{\rm tr}(\bb{\eta}) = 0$ and
$k^4\det(\bb{\eta}) = c^2k^2(\kdotb)^2/(4\pi e\nem)^2$, the WKB
dispersion relation (see \S \ref{S_wkb}) may immediately be
written down:
\begin{equation}\label{E_halldisp}
\left[\sigma^2 + i\,\sigma\frac{ck(\kdotb)}{4\pi e n_{\rm e}} +
\kdotva\right]\left[\sigma^2 -i\,\sigma\frac{ck(\kdotb)}{4\pi e
n_{\rm e}} + \kdotva\right] = -
2A\,\hall\bigl[\sigma^2+\kdotva\bigr]\,.
\end{equation}
This is identical to equation (81) in \citet{bt01} in the limit of
vanishing rotation frequency $\Omega$ (and negligible Ohmic
dissipation). Note that it is not necessary to assume $B_y/B_x\ll
1$, as in \S \ref{S_wkb}, since the only time-dependence is due to
the wavenumber $k$ (recall that the combination $\kdotb$ is a
constant).

Before we proceed any further in analyzing Equation
(\ref{E_halldisp}), it pays to examine the simple case of
vanishing shear. The dispersion relation then becomes
\begin{equation}
\sigma^2 \pm i\,\sigma \frac{ck(\kdotb)}{4\pi e\nem} + \kdotva =
0\,.
\end{equation}
This is precisely the dispersion relation for a uniformly
magnetized plasma with the displacement current and electron
inertia ignored (see, e.g., Krall \& Trivelpiece 1973); its
positive frequency solutions are
\begin{equation}\label{E_whistlers}
\omega = -i\sigma = \mp \,\frac{ck(\kdotb)}{8\pi e\nem} +
\left[\kdotva + \frac{c^2k^2(\kdotb)^2}{64\pi^2
e^2\nem^2}\right]^{1/2}\,.
\end{equation}
For small wavenumbers (low frequencies), both of the above
solutions reduce to Alfv\'{e}n waves. At larger wavenumbers,
however, right-handed waves (plus sign) go over to the
high-frequency whistler wave branch, whereas large $k$ left-handed
waves (minus sign) are cut off at a frequency
\begin{equation}
\omega_{\rm cutoff} = \frac{eB}{\mu c}\left(\frac{\nem}{n}\right)
\equiv \omega_{\rm c\mu}\left(\frac{\nem}{n}\right)\,,
\end{equation}
where $\mu$ is the mean mass per neutral particle. These modes
arise due to the drift of the field lines with respect to the ion
fluid as an Alfv\'{e}n wave propagates through the lighter
electron fluid, and are related to the so-called R- and L-waves of
plasma physics.

Restoring the velocity shear, it is straightforward to show that
instability proceeds through the point $\sigma=0$. The instability
criterion is therefore
\begin{equation}
\kdotva + 2A\,\hall < 0\,.
\end{equation}
This may be written in a more physically transparent form:
\begin{equation}\label{E_hallcrit}
\frac{2A}{\omega_{\rm c\mu}} < -\frac{\kdotb}{k_z
B}\left(\frac{\nem}{n}\right)\,.
\end{equation}
In other words, the time required for an ion to execute one
orbital gyration around a magnetic field line must be longer (by
at least the factor given on the right-hand side) than the time it
takes for a magnetic perturbation to grow by shear. If this
condition is not met, the ions are well-coupled to the electrons
(and thereby the magnetic field), and we are left with simple
linear-in-time growth due to shearing of the magnetic field
perturbation. The freedom in choosing the sign of $k_z(\kdotb)$
guarantees that any sign of shear can be destabilized, similar to
the result in \S \ref{S_case1} involving ambipolar diffusion.

\begin{figure}
\begin{center}
\leavevmode \qquad\qquad\qquad\qquad\quad
\includegraphics[width=3.27in]{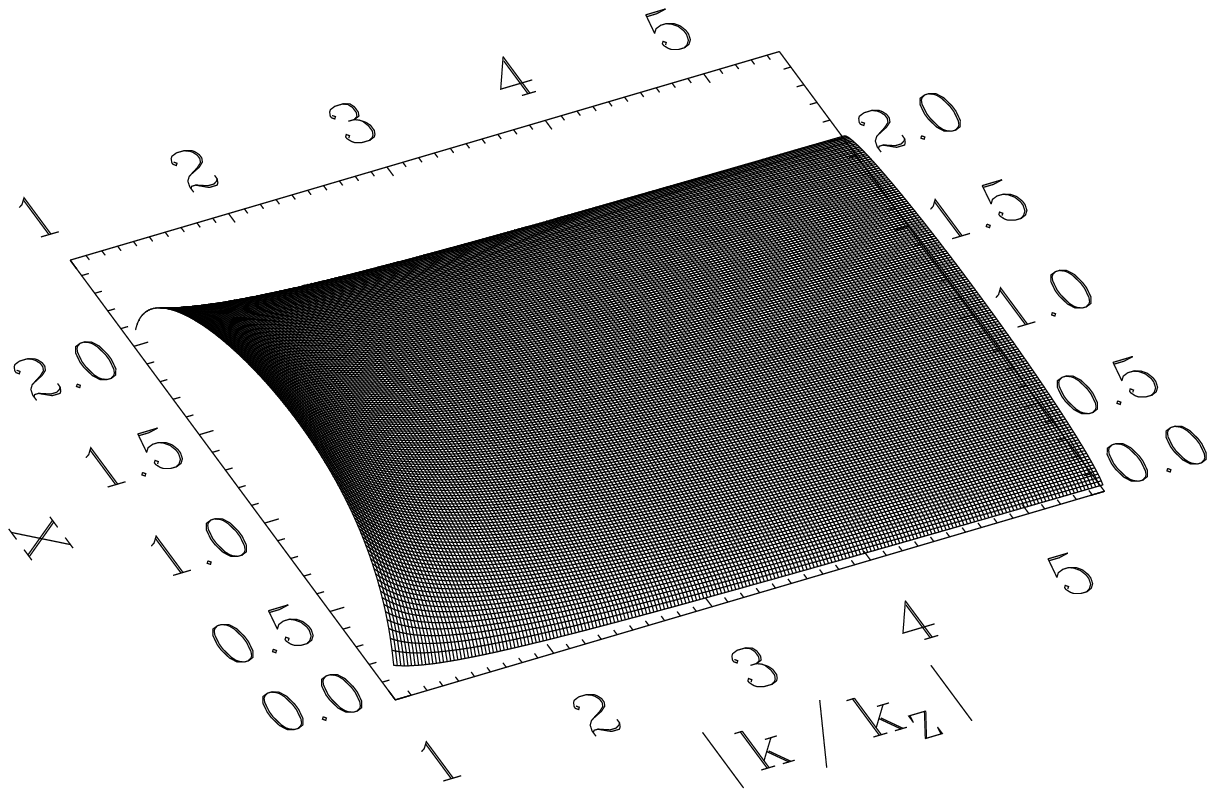}
\newline
\includegraphics[width=3.27in]{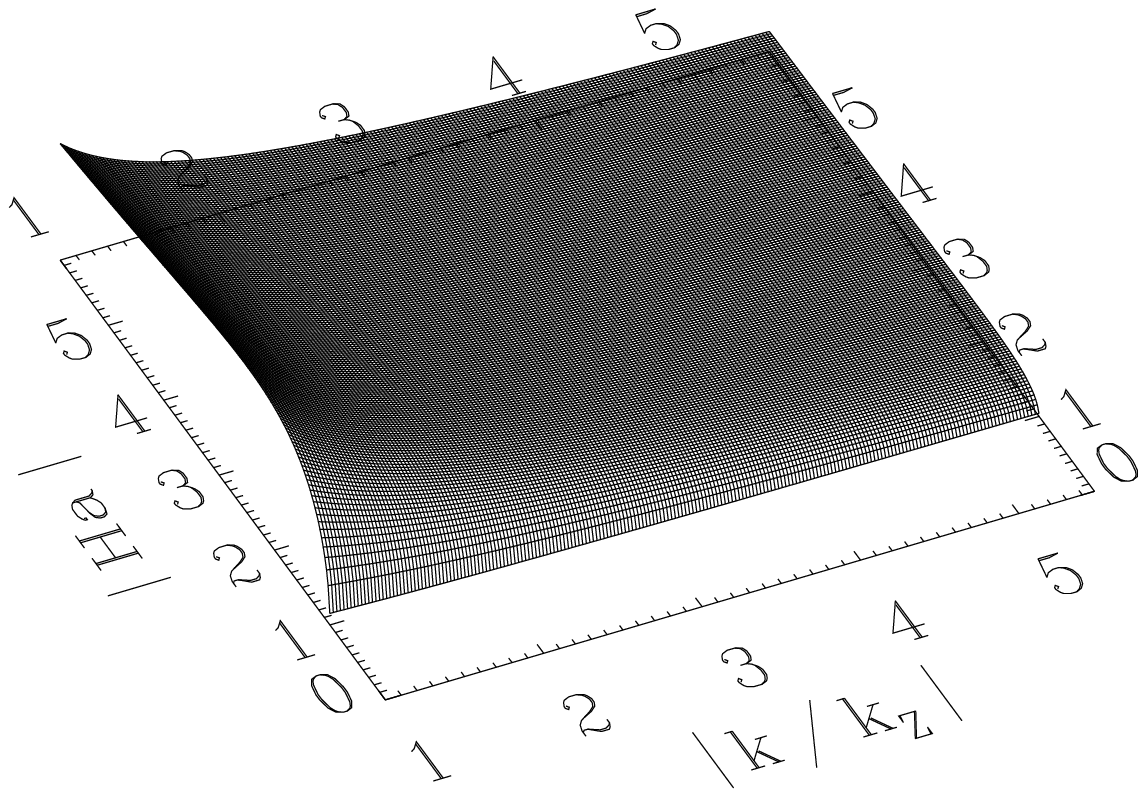}
\qquad
\includegraphics[width=3.27in]{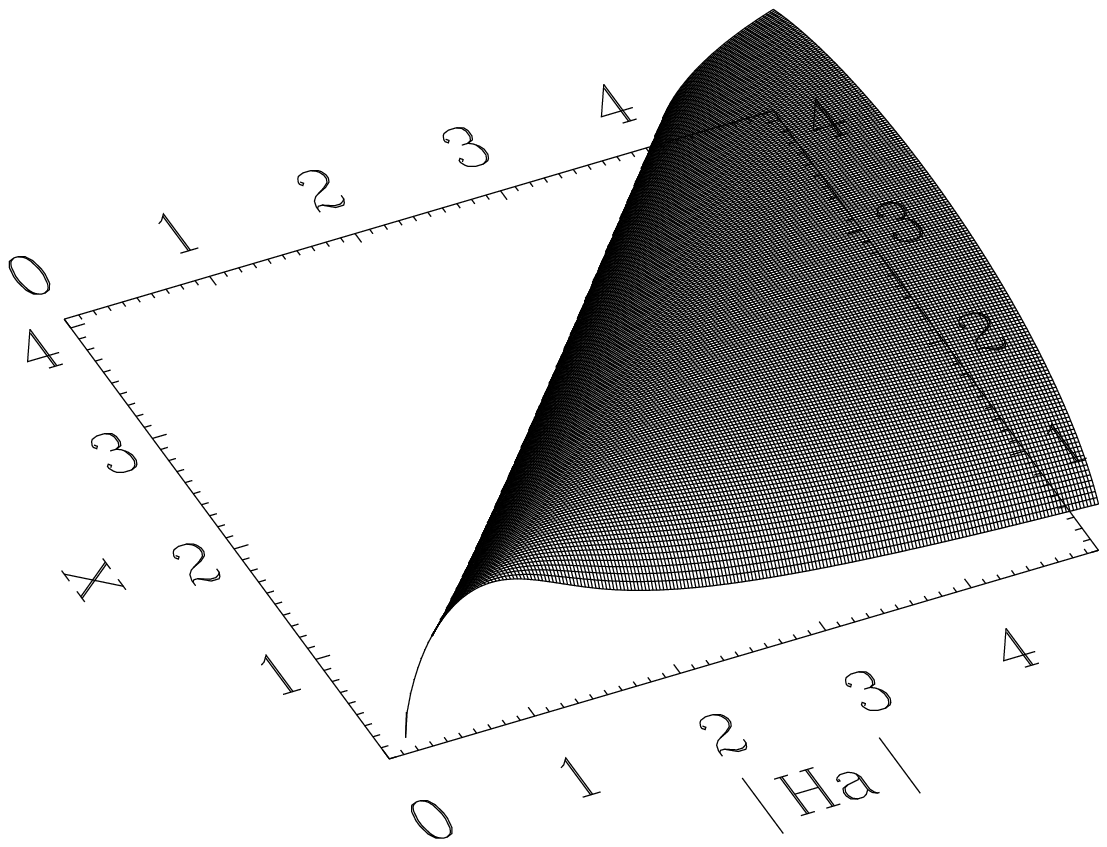}
\end{center}
\caption{{\it Counter-clockwise from top:} hall--shear instability
growth rates in (a) the $X$-$|k/k_z|$ plane with $|{\rm Ha}|=2$
and (b) the $|{\rm Ha}|$-$|k/k_z|$ plane with $X=1$; (c) maximum
growth in the $X$-$|{\rm Ha}|$ plane. Only regions of instability
are shown, with the height being proportional to the growth rate.
The maximum growth rate is $|A|$. } \label{F_hallgrowth}
\end{figure}

The maximum growth rate of this instability may be calculated
directly from Equation (\ref{E_maxgrowth}):
\begin{equation}
\sigma_{\rm max} = |A|\,,
\end{equation}
which occurs when $\bb{k}=k_z\ez$. This is in agreement with the
findings of \citet{bt01}, despite our neglect of rotation.
Defining the dimensionless parameters,
\begin{equation}
X \equiv \frac{\kdotva}{4A^2}\quad\quad{\rm and}\quad\quad {\rm
Ha} \equiv \frac{1}{|2A|}\hall\,,
\end{equation}
the dispersion relation may be written in dimensionless form and
growth rates may be determined numerically. In Fig.
\ref{F_hallgrowth}, we give three-dimensional plots of growth rate
in the $X$-$|k/k_z|$ plane (for $|{\rm Ha}|=2$) and in the $|{\rm
Ha}|$-$|k/k_z|$ plane (for $X=1$). The signs of $k_z(\kdotb)$ and
$2A$ are chosen such that instability is possible, and $k_y$ is
set to zero. Note that there is less unstable parameter space as
one goes to small $|{\rm Ha}|$ (the ion-neutral fluid becomes
well-coupled to the magnetic field). The maximum growth rate is
shown in Fig. \ref{F_hallgrowth}c for the parameter space spanned
by $X$ and $|{\rm Ha}|$. It is important to notice that if a
strongly-leading (or -trailing) wavevector ($|k/k_z| \gg 1$)
begins its life in the unstable regime, it will always be
unstable. By contrast, a stable wavevector will remain stable.

We had argued earlier (see \S \ref{S_behaviour}) that, in the
presence of shear, any physical mechanism that conspires to rotate
$\delta B_y$ back into $\delta B_x$ completes a feedback loop and
results in growth. HEMFs accomplish this task not by preferential
current damping, as in the case of the ambipolar-diffusion--shear
instability, but rather by current generation. Consider the case
of a purely vertical ($z$) wavenumber and magnetic field. The
effect of shear on the background magnetic field and the
coordinate system vanishes, and we may write down the linearized
induction Equations (\ref{E_bx}) and (\ref{E_by}) in the form
\renewcommand{\theequation}{\arabic{equation}\alph{subequation}}
\setcounter{subequation}{1}
\begin{equation}\label{E_hallbx}
\frac{d\delta B_x}{dt} + \hall\,\delta B_y = i(\kdotb)\delta
v_x\,,
\end{equation}
\addtocounter{equation}{-1}\addtocounter{subequation}{1}
\begin{equation}\label{E_hallby}
\frac{d\delta B_y}{dt} - \left[\hall + 2A\right]\delta B_x =
i(\kdotb)\delta v_y\,.
\end{equation}
\renewcommand{\theequation}{\arabic{equation}}
We assume that instability is possible, i.e., that the Hall and
shear terms in Equation (\ref{E_hallby}) have opposite signs. It
is clear from Equation (\ref{E_hallby}) that shear uses $\delta
B_x$ to generate a $\delta B_y$. The Hall terms, on the other
hand, generate $\delta B_x$ at the expense of $\delta B_y$. This
effect is actually present in the absence of shear, and arises
because the $y$-component of the perturbed electron velocity
differs from the ion-neutral velocity (recall that ambipolar
diffusion is ignored here) by a term involving $\delta J_y (\equiv
k_z \delta B_x - k_x \delta B_z)$. The induced magnetic field is
sheared further, leading to runaway. This behaviour can be seen
differently by viewing the Hall terms as `Coriolis' terms in the
magnetic field equations. These give birth to magnetic
`epicycles', i.e., circularly-polarized electromagnetic waves.
When the sign of the rotation imparted by the shear is opposite to
that of the handedness of these waves, a struggle ensues over
control of the direction of the perturbed magnetic field vector.
For every bit of field line stretching in the $y$ direction due to
shear, Hall forces rotate this increased field back into the $x$
direction, only to be stretched further by shear. The compromise
of this struggle is an exponentially-growing instability.

This route to instability has been seen before in the
Hall-modified MRI \citep{wardle99,bt01}, where the role of shear
is played by the differential rotation of an accretion disc [$v(x)
= x\Omega(x)$, where $\Omega$ is the orbital frequency]. In that
case, however, the behaviour of the Hall terms is complicated by
rotation, which influences the magnetic epicycles by introducing a
sense of helicity (see section 3 of Balbus \& Terquem 2001 for a
discussion). The finding that HEMFs render an accretion disc
unstable for both inwardly- {\em and} outwardly-decreasing angular
velocity profiles does not depend on rotational kinematics.
Rather, it is a simple result of the influence of shear on the
propagation of circularly-polarized electromagnetic waves.

At this point a few questions naturally emerge. First, why does
the Hall--shear instability require a vertical $(z)$ wavenumber in
order to operate, whereas the ambipolar-diffusion--shear
instability does not? Simply put, the interaction between shear
and the Hall effect is strongest when the motions implied by the
shear lie in the same plane as the magnetic ``epicycles" induced
by the Hall effect. In other words, instability is maximized when
the vorticity $\bb{\Gamma}\equiv\del\btimes\bb{v}$, the wavevector
$k$, and the magnetic field $B$ all share a mutual axis.
Translated mathematically, this implies that the relevant Hall
term is $(\bb{k}\bcdot\bb{B})(\bb{k}\bcdot\bb{\Gamma})$; it must
be negative for destabilization. This is distinct from the
findings of \citet{wardle99} and \citet{bt01}, who found that the
relevant coupling parameter is, respectively,
$(\bb{\Omega}\bcdot\bb{B})$ or
$(\bb{k}\bcdot\bb{B})(\bb{k}\bcdot\bb{\Omega})$ where
$\bb{\Omega}$ is the rotation vector. This corroborates our
finding that the Hall instability is not a result of differential
rotation per se, but rather of any source of shear. Our coupling
parameter matches theirs when the vorticity shares the same axis
as the rotation, a situation typical for disc systems. Second, why
is the growth rate for the Hall--shear instability typically so
much greater than for the ambipolar-diffusion--shear instability?
Both processes involve extraction of free energy from the
background shear flow. The reason lies in the fact that the Hall
effect employs a conservative process (cyclotron gyrations) rather
than a dissipative process (ambipolar diffusion) to rotate $\delta
B_y$ back into $\delta B_x$. Not surprisingly, the difference in
the growth rates is related to the rate at which the gas is heated
due to ion-neutral friction.

\section{Discussion: Shears, Rotations, and Projections}\label{S_discussion}

Despite the impression one may get from the abundance of
mathematical manipulations in the preceding sections, the
instabilities themselves are actually quite simple. They are the
result of combinations of shears, rotations (in the case of the
Hall effect), and projections (in the case of ambipolar
diffusion). Much in the way that the MRI may be understood by two
orbiting masses connected by a spring \citep{bh92a}, there exists
an equally intuitive toy model that captures the essence of the
shear instabilities examined in this paper. Through simple matrix
multiplication, several main results can be recovered without
recourse to the lengthy mathematical manipulations undertaken in
the preceding sections. All we require is some linear algebra.

Consider the following abstract problem, in which a position
vector $|r_0\rangle = |x_0,y_0\rangle$ (in Dirac notation), having
the coordinates $[x_0,y_0]$ in a two-dimensional $x$-$y$ Cartesian
coordinate system, is subjected to various shear, rotation, and
projection operators given by
\begin{equation}
\msb{S} \equiv \left[\begin{array}{cc} 1 & 0 \\
                                       \varepsilon &  1 \\
                     \end{array}\right]\,,
\qquad
\msb{R} \equiv \left[\begin{array}{rr} \cos\theta & -\sin\theta \\
                                       \sin\theta &  \cos\theta \\
                     \end{array}\right]\,,
\qquad{\rm and}\qquad
\msb{P} \equiv \left[\begin{array}{cc} \sin^2\phi & -\sin\phi\cos\phi \\
                                       -\sin\phi\cos\phi & \cos^2\phi \\
                     \end{array}\right]\,,
\end{equation}
respectively. We denote by the column vector $|r_n\rangle =
|x_n,y_n\rangle$ the position vector $|r\rangle$ after $n$
transformations have been applied to $|r_0\rangle$. To be precise,
the combination $\msb{S}|r\rangle$ results in a shearing of
$|r\rangle$ along the $y$-axis by a distance $\varepsilon x$;
$\msb{R}|r\rangle$ results in a counter-clockwise rotation of
$|r\rangle$ through an angle $\theta$; and, $\msb{P}|r\rangle$
projects $|r\rangle$ onto the unit vector
$|-\sin\phi,\cos\phi\rangle$. These will be put in a physical
context below.

\begin{figure}
\begin{center}
\includegraphics[angle=-90,width=6.9in,clip]{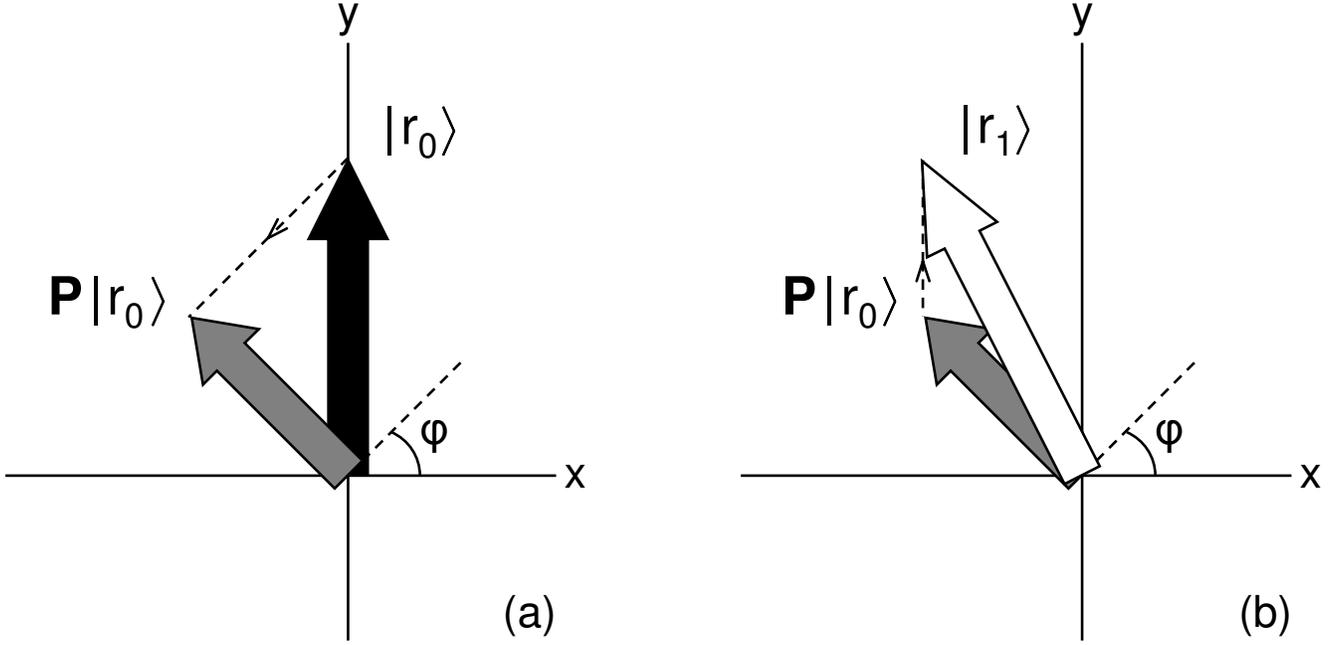}
\end{center}
\caption{Evolution of the initial state vector $|r_0\rangle$
(solid arrow) under the transformation $\msb{SP}$. (a) A
projection is applied to $|r_0\rangle$ that retains only its
component lying along the unit normal
$|-\sin\phi,\cos\phi\rangle$. The result is $\msb{P}|r_0\rangle$
(grey arrow). (b) This vector is then sheared along the $y$-axis
into $|r_1\rangle$ (open arrow). This process represents the
ambipolar-diffusion--shear instability (see text for details).
Figure is drawn to scale for comparison with Fig.
\ref{F_hallpicture}.}\label{F_ambipicture}
\end{figure}

We first turn our attention to the shear and projection operators.
Applying the combination $\msb{S}\msb{P}$ (a projection followed
by shear) to the initial state vector $|r_0\rangle$ advances it to
$|r_1\rangle$: \footnote{This requires
$|r_0\rangle\ne|\cos\phi,\sin\phi\rangle$, which corresponds to a
zero eigenvalue of the operator $\msb{SP}$. In this case, the
projection gives the null vector, and there is nothing left to
shear. For this particular initial state vector, the combination
$\msb{PS}$ works fine.}
\begin{equation}
| r_1\rangle = \msb{SP}|r_0\rangle
= (x_0\sin\phi - y_0\cos\phi)\left[\begin{array}{c} \sin\phi\\
                         \varepsilon\sin\phi-\cos\phi\\
       \end{array}\right]\,.
\end{equation}
A graphical depiction of this process is given in Fig.
\ref{F_ambipicture}. While $|r_0\rangle$ is clearly not an
eigenvector of the operator $\msb{SP}$, it is straightforward to
show that $|r_1\rangle$ {\em is} an eigenvector, with eigenvalue
$(1-\varepsilon\sin\phi\cos\phi)$. It is then trivial to write
down the result for general $n \ge 1$:
\begin{equation}\label{E_ambirecursive} | r_n\rangle =
(1-\varepsilon\sin\phi\cos\phi)\,|r_{n-1}\rangle =
(1-\varepsilon\sin\phi\cos\phi)^{n-1}\,| r_1\rangle \,.
\end{equation}
Put differently, after this transformation has been performed just
once, all subsequent applications simply stretch (or contract) the
vector along $|r_1\rangle$ by a factor equal to the associated
eigenvalue. Note that in order for any evolution to occur,
$\phi\ne m\pi/2$ with $m$ an integer. For growth, we require
\begin{equation}\label{E_projectioncriterion}
\varepsilon\sin\phi\cos\phi < 0\,.
\end{equation}
We now place these results in the context of the
ambipolar-diffusion--shear instability. With $\varepsilon =
2A\Delta t$ and $|r\rangle=\delta\bb{B}$, our shear operator
corresponds physically to the production of $\delta B_y$ due to
the shearing of $\delta B_x$ in a time $\Delta t$. When $\phi$ is
taken to be the angle between a background magnetic field and the
$x$-axis, the projection operator embodies ambipolar diffusion
acting on the vector $|r\rangle=\delta\bb{B}$. Their combination
leads precisely to the behaviour described in \S
\ref{S_behaviour}. In fact, it embodies the {\em exact} solution
for the case $\bb{k}=k_z\ez$ and $\bb{B} = B\cos\phi\ex +
B\sin\phi\ey$, as can be readily verified by taking the limit
$\Delta t\rightarrow 0$ in Equation (\ref{E_ambirecursive}) to
arrive at the differential equation:
\begin{equation}
\frac{dr}{dt} = -A\sin(2\phi)\,r\,.
\end{equation}
If $A\sin(2\phi)<0$, we have exponential growth. The growth rate
is $A\sin(2\phi)$, which has its maximum at precisely the Oort $A$
value. That the maximum growth rate occurs at $\phi=\pi/4$ is in
accord with the well-known ``$\sin\,2l$" law (e.g., Mihalas \&
Binney 1981), a result of particular significance; we refer the
reader to \S 2.4 of \citet{bh92a} for a full discussion of this
topic.

\begin{figure}
\begin{center}
\includegraphics[angle=-90,width=6.9in,clip]{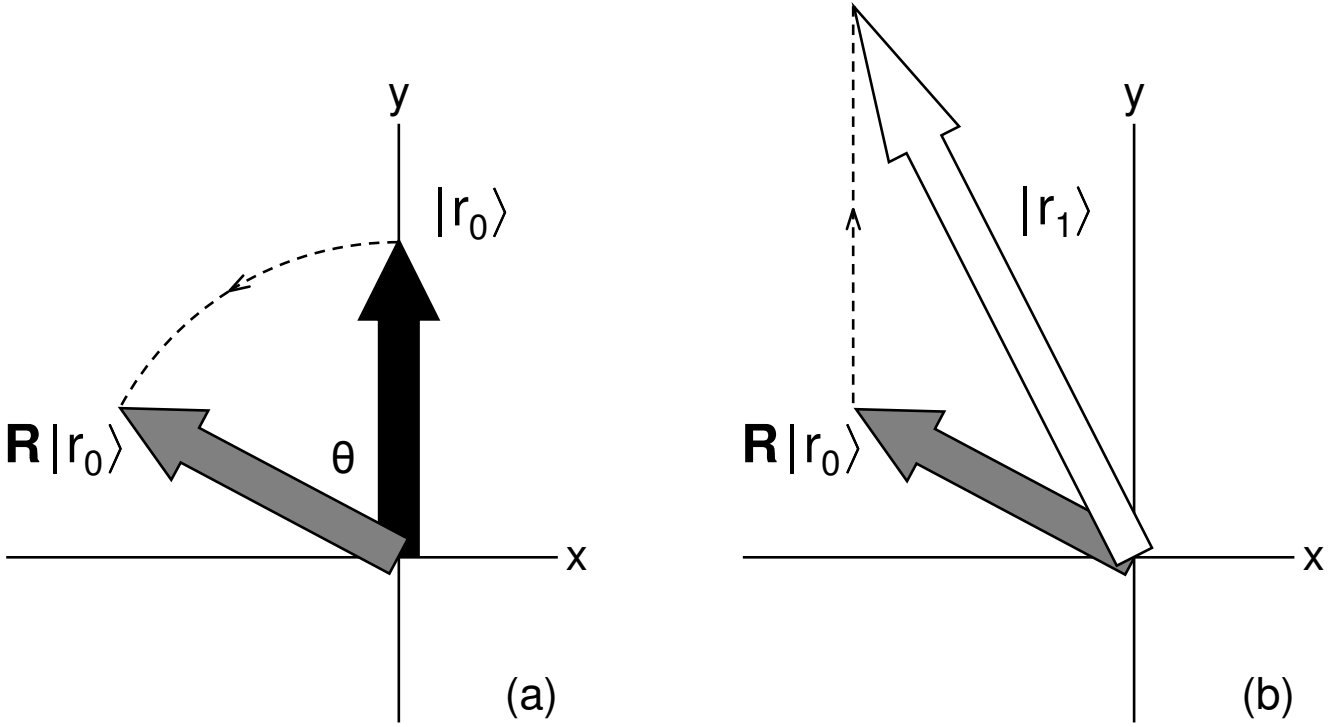}
\end{center}
\caption{Evolution of the initial state vector $|r_0\rangle$
(solid arrow) under the transformation $\msb{SR}$. (a) A
counter-clockwise rotation by $\theta$ is applied to
$|r_0\rangle$, taking it into $\msb{R}|r_0\rangle$ (grey arrow).
(b) This vector is then sheared along the $y$-axis into
$|r_1\rangle$ (open arrow). The process represents the Hall--shear
instabiliy (see text for details). Figure is drawn to scale for
comparison with Fig. \ref{F_ambipicture}.}\label{F_hallpicture}
\end{figure}

Next we turn our attention to the shear and rotation operators.
Applying the combination $\msb{S}\msb{R}$ (a rotation followed by
shear) to the initial state vector $|r_0\rangle$ advances it to
$|r_1\rangle$:
\begin{equation}
| r_1\rangle = \msb{SR}|r_0\rangle
= (x_0\sin\theta - y_0\cos\theta)\left[\begin{array}{c} 1\\
                         \varepsilon\\
       \end{array}\right] + (x_0\sin\theta + y_0\cos\theta)\left[\begin{array}{c} 0\\ 1\\ \end{array}\right]\,.
\end{equation}
A graphical depiction of this process is given in Fig.
\ref{F_hallpicture}. For $n \ge 2$, it is possible to show that a
recursion relation exists between successive state vectors:
\begin{equation}\label{E_hallrecursive}
|r_n\rangle - (2\cos\theta-\varepsilon\sin\theta)|r_{n-1}\rangle +
|r_{n-2}\rangle = 0\,.
\end{equation}
This equation has the general solution:
\begin{equation}
|r_n\rangle = \left(\frac{\lambda^n_+ - \lambda^n_-}{\lambda_+ -
\lambda_-}\right)\,|r_1\rangle - \left(\frac{\lambda^{n-1}_+ -
\lambda^{n-1}_-}{\lambda_+ - \lambda_-}\right)\,|r_0\rangle\,,
\end{equation}
where
\begin{equation}
\lambda_\pm = \left(\cos\theta -
\frac{\varepsilon}{2}\sin\theta\right) \pm \left[\left(\cos\theta
- \frac{\varepsilon}{2}\sin\theta\right)^2 - 1\right]^{1/2}
\end{equation}
are the characteristic roots of Equation (\ref{E_hallrecursive}).
With $\theta = \omega\Delta t$ and $|r\rangle=\delta\bb{B}$, this
transformation corresponds to the physical behaviour described in
\S \ref{S_hall} concerning the evolution circularly-polarized
electromagnetic waves (with frequency $\omega$) in the presence of
shear; namely, in a time $\Delta t$, the perturbed magnetic field
vector is rotated by $\msb{R}$ through an angle $\omega\Delta t$
and sheared by $\msb{S}$ along the $y$-axis. Taking the $\Delta t
\rightarrow 0$ limit of Equation (\ref{E_hallrecursive}) gives the
differential equation
\begin{equation}
\frac{d^2 r}{dt^2} = -\omega(\omega+2A)\,r\,,
\end{equation}
whose solutions are exponentially growing if
\begin{equation}
\frac{2A}{\omega} < -1\,.
\end{equation}
The similarity between this instability criterion and Equation
(\ref{E_hallcrit}) is striking. When $|A/\omega|$ equates to
unity, the growth rate attains its maximum value: precisely the
Oort $A$ constant. This simple model captures all the salient
features of the Hall--shear instability.

\section{Summary}\label{S_summary}

In this paper, we have investigated the stability of
weakly-ionized, magnetized planar shear flows to linear
disturbances. Employing a local approach similar to the
shearing-sheet approximation of \citet{glb65}, we have derived two
coupled differential equations governing the evolution of magnetic
field perturbations in the presence of either ambipolar diffusion
or the Hall effect. Solutions are found by WKB methods and by
direct numerical integration. We find that instability arises from
the combination of shear and non-ideal MHD processes, and is a
result of the ability of these processes to open new pathways for
the fluid to feed off the free energy of shear. They turn what
would be simple linear-in-time growth due to current and vortex
stretching from shear into exponential instabilities. We have also
constructed a simple toy model based on transformation operators
that not only captures all the qualitative results of this paper,
but also matches the exact {\em quantitative} solution in some
specific instances.

In the case of ambipolar diffusion, anisotropic damping leads to
the generation of magnetic field perturbations perpendicular to
the background magnetic field. What ensues is a competition
between ambipolar diffusion and shear, which is trying desperately
to stretch the magnetic field along the stream-wise direction. In
the end they both win, and the perturbations grow in a direction
somewhere between the two that depends upon the geometry of the
magnetic field and the ratio of the neutral-ion collision
time-scale to the shearing time-scale. The resulting growth rates
are on the order of $0.1\,|2A|$ (in the case of a time-independent
background). It is notable that, in the general case of a
time-dependent background, the growth rates increase in time
without bound since highly-trailing (or -leading) shearing waves
are trivially unstable. If the exponential growth phase lasts
sufficiently long, growth rates may become comparable to or even
exceed the shearing rate.

In the case of the Hall effect, it is not current damping but
rather current generation that gives rise to a magnetic field
component perpendicular to the shear. Instability arises from the
influence of shear on the propagation of circularly-polarized
electromagnetic (whistler) waves, and is present so long as the
shearing frequency $|2A|$ is larger than the ion cyclotron
frequency (times a factor proportional to the degree of
ionization). Growth rates are on the order of $0.5\,|2A|$. In
contrast to the work of \citet{wardle99} and \citet{bt01}, we find
that instability depends not on $\bb{\Omega}\bcdot\bb{B}$ or
$(\bb{k}\bcdot\bb{B})(\bb{k}\bcdot\bb{\Omega})$, respectively, but
rather on $(\bb{k}\bcdot\bb{B})(\bb{k}\bcdot\bb{\Gamma})$, where
$\Gamma=\del\btimes\bb{v}$ is the vorticity of the background.
This should be negative for destabilization. Provided a given
wavevector begins its evolution unstable (stable), it will remain
unstable (stable) until non-linear processes intercede. The
physical reason that typical growth rates for the Hall--shear
instability are so much greater than the maximum growth rate for
the ambipolar-diffusion--shear instability is that the Hall--shear
instability employs a conservative process (cyclotron gyrations)
rather than a dissipative process (ambipolar diffusion). The
difference in these rates is related to the rate at which
ion-neutral friction heats the gas.

In both cases, unstable wavenumbers can be found for any sign of
the velocity gradient. This explains why these processes were
found to destabilize both inwardly- and outwardly-decreasing
angular velocity gradients in accretion disks \citep{bt01,kb04}.
While the analysis is complicated somewhat by the time-dependence
of the shearing background, we find that the evolution unfolds as
a series of time-independent problems, similar to studies of
nonaxisymmetric instabilities in discs (e.g., Balbus \& Hawley
1992b, Balbus \& Terquem 2001). In the limit of small horizontal
($y$) wavenumber, $k_y/k_z \ll 1$, the instability criterion may
be written
\begin{equation}
\kdotva + k^2\bb{\eta}\,\bb{:}\,\del\,\bb{v} < 0\,,
\end{equation}
with the resistivity tensor $\bb{\eta}$ being given by either
Equation (\ref{E_eta}) (for ambipolar diffusion) or Equation
(\ref{E_eta2}) (for the Hall effect). The final term here is the
key to instability; in formal Cartesian index notation $(i,j,k)$,
it is $k^2\eta_{ij}\partial v_j/\partial x_i$. The maximum growth
rate for a given $\bb{\eta}$ is
\begin{equation}
\sigma_{\rm max} =
\left|\frac{\bb{\eta}\,\bb{:}\,\del\,\bb{v}}{2\det^{1/2}(\bb{\eta})
+\,{\rm tr}(\bb{\eta})}\right|\,.
\end{equation}
For both non-ideal MHD effects, this is independent of the degree
of ionization. Off-diagonal elements in the resistivity tensor are
essential. No shear instabilities are present for isotropic
damping processes, such as Ohmic dissipation.

The impact of these results on astrophysical systems is unclear at
this point. Our neglect of rotation precludes a straightforward
application to accretion discs. There is a fundamental difference
between planar shear flows and disc systems: in a planar shear
flow there is only one characteristic gradient, $dv/dx$, whereas
in a disk system there are two, one for the angular velocity, and
one for the angular momentum. In astrophysical discs, the omitted
Coriolis force generally dominates the shear dynamics. There is no
asymptotic domain for either linear or non-linear perturbations in
which the governing dynamical equations behave locally like
Cartesian shear. It seems that the best we can do here is to claim
linear instability in constant-specific-angular-momentum discs. In
addition, the well-known fact that inviscid planar shear layers
are hosts to a plethora of non-linear instabilities suggests that
the non-ideal MHD effects investigated in this paper may play only
secondary roles.

This may be undue pessimism, however. The instabilities
investigated in this paper {\em do} have rotational counterparts,
even in Keplerian systems. The real utility of this calculation,
therefore, is that by removing rotation from the problem we obtain
a clearer physical picture of what is going on in actual discs.
One important simplification is that the MRI is absent. This is
expected, of course, since the MRI is not {\em just} a shear
instability. Rather, the MRI plays the crucial role of
redistributing angular momentum and thereby opening paths to lower
energy states in differentially-rotating discs. On the other hand,
the ambipolar-diffusion-- and Hall--shear instabilities have no
preference for the source of the shear, whether it be in linear or
angular velocity. The extent to which they lead to enhanced
angular momentum transport is not governed by magnetic stresses,
as in the case of the MRI, but rather whether or not the bulk
fluid is responsive enough to utilize an increasingly radial
magnetic field.

\section*{Acknowledgments}

This work has benefited from enlightening conversations with Bryan
Johnson, whose detailed comments on an early draft of this paper
led to a much improved presentation, and Steve Balbus, whose
suggestion that the work in \S \ref{S_AD} on ambipolar diffusion
might carry over to the Hall effect led to \S \ref{S_hall}. I also
thank Steve Desch, Telemachos Mouschovias, and Konstantinos Tassis
for useful discussions, and an anonymous referee for constructive
comments following a careful reading of the manuscript.

\bsp\label{lastpage}
\end{document}